\documentclass[aoas,MSNbibl,nameyear,seceqn,dvips]{arximspdf}
\usepackage{mathrsfs,mathbh,dcolumn}
\usepackage{framed}
\usepackage{graphicx}
\usepackage{url,breakurl}

\doi{10.1214/14-AOAS760} 
\volume{8}
\issue{3}
\pubyear{2014}
\firstpage{1853}
\lastpage{1891}
\docsubty{FLA}

\makeatletter
\def\bar{\overline}
\newcolumntype{d}[1]{D{.}{.}{#1}}
\newcommand{\eqref}[1]{(\ref{#1})}
\newtheorem{theorem}{Theorem}

\newproclaim{definition}[theorem]{Definition}
\newproclaim{observation}[theorem]{Observation}
\newproclaim{remark}[theorem]{Remark}
\newproclaim{problem}{Problem}
\newcommand{\Cal}{\mathcal}
\newcommand{\weakc}{\stackrel{{w}}{\longrightarrow}}

\newcommand{\vd}{\mathbf{d}}
\newcommand{\vp}{\mathbf{p}}
\newcommand{\vq}{\mathbf{q}}
\newcommand{\bN}{\mathbb{N}}
\newcommand{\bR}{\mathbb{R}}
\newcommand{\erdos}{Erd\H{o}s--R\'enyi }
\makeatother

\begin{document}
\begin{frontmatter}

\title{A testing based extraction algorithm for identifying significant
communities in networks\thanksref{T1}}
\runtitle{Identifying significant communities in networks}
\thankstext{T1}{Supported in part by NSF Grants DMS-09-07177,
DMS-13-10002, DMS-06-45369, DMS-11-05581 and SES-1357622.}

\begin{aug}
\author[A]{\fnms{James~D.}~\snm{Wilson}\corref{}\ead[label=e1]{jameswd@email.unc.edu}},
\author[B]{\fnms{Simi}~\snm{Wang}\ead[label=e2]{wangsimi@email.unc.edu}},
\author[C]{\fnms{Peter~J.}~\snm{Mucha}\ead[label=e3]{mucha@email.unc.edu}\thanksref{T2}},
\author[A]{\fnms{Shankar}~\snm{Bhamidi}\ead[label=e4]{bhamidi@email.unc.edu}}
\and
\author[A]{\fnms{Andrew~B.}~\snm{Nobel}\ead[label=e5]{nobel@email.unc.edu}}
\runauthor{Wilson et al.}
\affiliation{University of North Carolina at Chapel Hill}
\thankstext{T2}{Supported in part by the James S. McDonnell Foundation
21st Century Science Initiative---Complex Systems Scholar Award Grant
220020315.}
\address[A]{J.~D. Wilson\\
S. Bhamidi\\
A.~B. Nobel\\
Department of Statistics \\
\quad and Operations Research\\
University of North Carolina\\
\quad at Chapel Hill\\
Chapel Hill, North Carolina 27599\\
USA\\
\printead{e1}\\
\phantom{E-mail:\ }\printead*{e4}\\
\phantom{E-mail:\ }\printead*{e5}}
\address[B]{S. Wang\\
Department of Mathematics\\
University of North Carolina at Chapel Hill\\
Chapel Hill, North Carolina 25799\\
USA\\
\printead{e2}}
\address[C]{P.~J. Mucha\\
Department of Applied Physical Sciences\\
University of North Carolina\\
\quad at Chapel Hill\\
Chapel Hill, North Carolina 25799\\
USA\\
\printead{e3}}
\end{aug}
%

\received{\smonth{9} \syear{2013}}
\revised{\smonth{5} \syear{2014}}

%
\begin{abstract}
A common and important problem arising in the study of networks is how
to divide the vertices of a given network into one or more groups,
called communities, in such a way that vertices of the same community
are more interconnected than vertices belonging to different ones.
We propose and investigate a testing based community detection
procedure called Extraction of Statistically Significant Communities (ESSC).
The ESSC procedure is based on $p$-values for the strength of connection
between a single vertex and a set of vertices under a reference
distribution derived from a conditional configuration network model.
The procedure automatically
selects both the number of communities in the network and their size.
Moreover, ESSC can handle overlapping communities and, unlike the
majority of
existing methods, identifies ``background'' vertices that do not belong
to a well-defined community. The method has only one parameter, which controls
the stringency of the hypothesis tests.
We investigate the performance and potential use of ESSC and compare it
with a number of existing methods, through a validation study using
four real network data sets. In addition, we carry out a simulation
study to assess the effectiveness of ESSC in networks with various
types of community structure, including networks with overlapping
communities and those with background vertices. These results suggest
that ESSC is an effective exploratory tool for the discovery of
relevant community structure in complex network systems. Data and
software are available at \url{http://www.unc.edu/\textasciitilde
jameswd/research.html}.
\end{abstract}

%
\end{frontmatter}

\section{Introduction}\label{sec:intro}
\begin{figure}[b]

\includegraphics{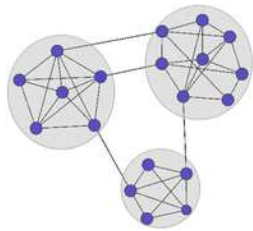}

\caption{A simple network with three distinct communities.}\label
{fig:stoch3}
\end{figure}

The study of networks has been motivated by, and made
significant contributions to, the modeling and understanding of complex
systems. Networks are used to model the relational structure between
individual units of an observed system. In the network setting,
vertices represent the units of the system and edges are placed between
vertices that are related in some way. Network-based models have been
used in a variety of disciplines: in biology to model protein-protein
and gene--gene interactions; in sociology to model friendship and
information flow among a group of individuals; and in neuroscience to
model the relationship between the organization and function of the
brain. In many of these applications, the vertices of the network under
study can naturally be subdivided into communities. Informally, a
community is a group of vertices that are more connected to each other
than they are to the remainder of the network. More rigorous
definitions quantify this notion of differential connection in
different ways.
Figure~\ref{fig:stoch3} illustrates a network with three disjoint communities.
%
%
%

The problem of dividing the vertices of a given network into
well-defined communities is known as community detection.
Community detection has become increasingly popular, as communities
have been found to identify important
and useful features of many complex systems. Community detection has
been studied by researchers in a variety of fields, including
statistics, the social sciences,
computer science, physics and applied mathematics, and a diverse set of
community detection algorithms have been developed [see \citet
{fortunato2010community,porter2009communities} for reviews].

Existing community detection methods capture different types of
community structure. The simplest community structure, and the one most
commonly studied,
is a hard partitioning, in which each vertex of the network is assigned
to one and only one community,
and the collection of communities together form a partition of the
network [e.g., \citet{newman2004finding,ng2002spectral,snijders1997estimation}].
Another class of community structure allows overlapping communities
[see \citet{xie2011overlapping} for a recent review], in which
the collection of communities together form a cover of the network.
Broadly speaking, most community detection methods produce one of these
types of structures.

Community detection has been successful in understanding a wide variety
of complex systems. In addition to the numerous examples cited in the
aforementioned reviews, community detection methods have recently been
profitably applied to
protein interaction networks [\citet{Lewis2010}],
functional brain activity [\citet{Bassett2011}],
social media [\citet{Papadopoulos2012}] and
mobile phone data [\citet{Muhammad2013}], as well as
social groups [\citet{Greene2010,Miritello2011,Onnela2011}].

The majority of existing community detection methods make the
assumption that \textit{every} vertex within an observed network belongs
to at least one community.
Though many networks can be appropriately divided into a partition (or
cover) of communities, some large and heterogeneous networks do not fit
into this framework. {For example, consider the Enron email network
from \citet{leskovecENRON} where edges represent the email
correspondence (sent or received) between email accounts in 2001. The
network contains many (on the order of 10K) email accounts outside of
Enron and relatively few (on the order of 1K) email accounts from
employees at Enron. The outside email accounts, many of which are spam
email accounts, are not preferentially attached to any group of
employees and thereby do not belong to a well-defined community.} From
this example and several others that we investigate in Section~\ref
{sec:realdata}, we will see that many real networks contain vertices
that do not have strong connections to \textit{any} community. Informally,
we call vertices that are not preferentially connected to any community
\textit{background} vertices, as they act as a background against which
more standard community structures may be detected.

\begin{figure}

\includegraphics{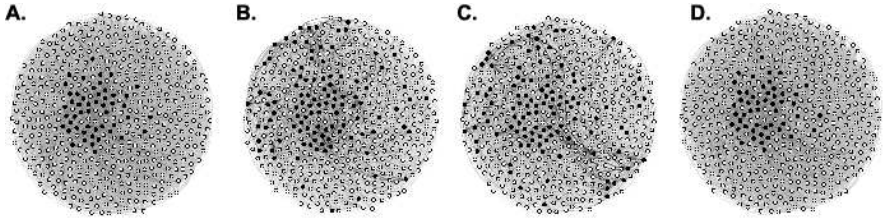}

\caption{\textup{(A)} A toy network that contains one significantly
connected community---colored in black---and many sparsely connected
background vertices. \textup{(B)} The partition given by the GenLouvain
modularity optimization method. \textup{(C)} The partition given by
normalized Spectral clustering. \textup{(D)}~The extracted community found
by the proposed method ESSC, which separates and distinguishes the
embedded community from the background.} \label{fig:toy}
\end{figure}

In networks where background vertices are present, partitioning and
covering methods
typically assign them to more tightly connected communities. To
illustrate this, we generated a 500 node toy network with a single
community of size~50,
whose vertices are linked independently with probability 0.5; the
remaining vertices are background and are linked to all vertices in the
network independently with probability 0.05. We ran two popular
detection methods---the modularity based algorithm of \citet
{newman2004finding} and the normalized Spectral algorithm of \citet
{ng2002spectral}---and found two disjoint communities. We considered
the community that most closely matched the true embedded community and
found, as shown in Figure~\ref{fig:toy}, that both methods included
many background vertices.

Also shown in Figure~\ref{fig:toy} is the result of applying the ESSC
method introduced in this paper. ESSC accurately identifies the
embedded community and the background, and separates one from the other.
Although there are methods in multivariate clustering to capture
background [\citet{ester1996density,hinneburg1998efficient}],
only a few recent papers, for example, \citet
{zhao2011community,lancichinetti2011finding}, consider background in
the context of
community detection.

%
%

%

In this paper we propose and study a testing based community detection
algorithm, called
Extraction of Statistically Significant Communities (ESSC), that is
capable of identifying both
background vertices and overlapping communities. The core of the
algorithm is an iterative
search procedure that identifies statistically stable communities. In
particular, the search procedure uses
tail probabilities derived from a stochastic configuration model based
on the observed network
in order to assess the strength of the connection between a single
vertex and a candidate community.
Updating of the candidate community is carried out using
ideas from multiple testing and false discovery rate control.

The only free parameter in the ESSC algorithm is a false discovery rate
threshold that is used
in the update step of the iterative search procedure. The number of
detected communities, their overlap
(if any) and the size of the background are handled automatically,
without user input. In practice,
the output of ESSC is not overly sensitive to the threshold parameter;
see the Appendix~\ref{sec:alpha} for more details.
\subsection{Notation}\label{sec:notation}
For ease of discussion throughout the remainder of this paper, we first
introduce some notation. Let $G = (V,E)$ be an {undirected}
{multigraph} with vertex set $V = [n] = \{1,\ldots,n\}$ and edge
{multiset} $E$ containing all (unordered) pairs $\{i,j\}$ such
that there is an edge between vertices $i$ and $j$ in $G$, allowing
repetitions for multiple edges. Let $d(u)$ denote the degree of a
vertex $u$,
and let $\mathbf{d} = \{ d(1), \ldots, d(n) \}$ denote the degree sequence
of $G$. Let $B \subset[n]$ denote a subset of vertices in $G$. Indices
on $B$ are simply used for specification throughout. Write $\Pi$ for a
partition of the vertex set $[n]$ ($\Pi= B_1 \cup B_2 \cup\cdots\cup
B_k$, $k \geq1$). In many cases, detection methods seek a partition
(or cover) through optimizing a specified quality or score function,
which we will denote as $S(\cdot)$. It is important to note that the
score may be global, in which case $S(\cdot)$ measures the quality of
an entire partition, or local, in which case $S(\cdot)$ measures the
quality of a potential community. We will use $G_o$ to denote an
observed graph and $\widehat{G}$ for a stochastic model on the vertex
set~$[n]$.
\subsection{Related work}\label{sec:relatedwork}
There is an extensive literature on the development and analysis of
community detection methods. In this section we give an overview of
this literature. For recent surveys describing community detection
methods, see \citet{fortunato2010community,porter2009communities} or
\citet{ goldenberg2010survey}. In Section~\ref{sec:competingmethods} we
describe in more detail the methods to which we compare ESSC.

Many of the earliest community detection methods approach network
clustering from a graph-theoretic standpoint. Relying on a prespecified
integer $k$, these methods seek the partition of $k$ communities that
minimize the number of edges between communities. The optimal partition
of this criterion is known as the partition of min-cut and max-flow
[\citet{goldberg1988new}], where the \textit{cut} of a community specifies
the number of edges from the community to the rest of the network.
Unfortunately, min-cut methods often result in many singleton
communities. To address this issue, the cut of a community can be
normalized by either the community size, resulting in the ratio-cut
criterion [\citet{wei1989towards}], or by the total degree of the
community, giving the normalized-cut criterion [\citet
{shi2000normalized}]. When $k > 2$, the task of finding the partition
that satisfies any of these cut criterions is NP-hard. Spectral
clustering methods [\citet{krzakala2013spectral,ng2002spectral}] find an
approximate solution to the norm-cut criterion by appealing to spectral
properties of the graph Laplacian. Spectral clustering methods can be
applied to either nonnetwork multivariate data or directly to
relational network data.

Another class of community detection methods seek community structure
by comparing the observed network $G_o = ([n],E_o)$ with an
unstructured stochastic network on the same vertex set $\widehat
{G}_{\mathrm{null}} = ([n], \hat{E}_{\mathrm{null}})$. A stochastic network $\widehat
{G}_{\mathrm{null}}$ describes the probabilities of edge connection between all
pairs of vertices in $[n]$ given that each pair was connected at
random. Detection methods of this class seek the partition of $G_o$
whose clustering most deviates from what is expected under $\widehat
{G}_{\mathrm{null}}$. Modularity methods [see, e.g., \citet
{blondel2008fast,clauset2004finding,newman2006modularity,mucha2010community}]
are a
popular subset of this class. Modularity methods seek the partition
whose communities' fraction of observed edges are furthest from the
fraction of edges expected under $\widehat{G}_{\mathrm{null}}$, that is, the
partition $\Pi$ that maximizes
\[
\label{eq:mod} S_{\mathrm{mod}}(\Pi) = \frac{1}{2|E_o|} \sum
_{\ell= 1}^k \biggl(\sum_{i,j \in B_{\ell}}
\mathbb{I}\bigl(\{i,j\} \in E_o\bigr) - \gamma\mathbb{E} \biggl(\sum
_{i,j \in B_{\ell}} \mathbb{I}\bigl(\{ i,j\} \in
\hat{E}_{\mathrm{null}}\bigr) \biggr) \biggr),
\]
where $\gamma> 0$ is a resolution parameter that controls
the size of discovered communities. In many cases, $\gamma$ is treated
as one, however, this parameter can be tuned in a data-driven fashion.
There are many choices for a reference stochastic network. For
instance, in the case of the Newman--Girvan modularity [\citet
{newman2004finding}], $\widehat{G}_{\mathrm{null}}$ is specified as the
configuration model [\citet{molloy1995critical}] under which the degree
sequence of $G_o$ is maintained. In this case $\mathbb{E} (\sum_{i,j \in B_{\ell}} \mathbb{I}(\{i,j\} \in\hat{E}_{\mathrm{null}}) )$ is
$d_o(i) d_o(j) / 2|E_o|$. Our proposed method ESSC also relies upon the
configuration model as a reference stochastic network.

An alternative class of community detection methods estimate the
community structure of a network by fitting a structured stochastic
network $\widehat{G}_{\mathrm{struct}} = ([n], \hat{E}_{\mathrm{struct}})$ to the
observed data $G_o$. Here, $\widehat{G}_{\mathrm{struct}}$ describes random
assignments of edges conditional on stochastic community (or block)
structure on the vertex set~$[n]$. Formally, $\widehat{G}_{\mathrm{struct}}$
is a parametric model whose parameters describe the community labels of
each vertex and potentially the topological properties of the network
(e.g., the degree distribution of the network). Given an observed
network $G_o$ and a prespecified integer $k$, a structured network
(with parameters $\Theta$) is fit to $G_o$ by maximizing the likelihood
function describing $\Theta$: $\mathcal{L}(\Theta|G_o,k)$. A recent
review of structured network models is provided by \citet
{goldenberg2010survey}. One of the most popular network models of this
type is the stochastic block model [\citet
{holland1983stochastic,snijders1997estimation,nowicki2001estimation}].
Under this model,
vertices are assigned labels taking values in $\{1,\ldots,k\}$
according to probabilities ${\pi} = (\pi_1, \ldots,\pi_k)$. Conditional
on the vertex labels, edge probabilities are given by a $k \times k$
symmetric matrix $\mathbf{P}$ where the $i$, $j$th entry of $\mathbf{P}$
gives the probability of an edge between community $i$ and $j$. Block
models are fit to $G_o$ by maximizing the corresponding likelihood
$\mathcal{L}(\Theta= (\mathbf{P},\pi)|G_o,k)$. Other examples of
structured stochastic networks include latent variable models [\citet
{hoff2002latent,handcock2007model}] and mixed membership models which
are flexible to overlapping communities [\citet
{airoldi2008mixed,ball2011efficient}].

{Recently, there has been significant progress in the development of
fast and efficient algorithms for fitting stochastic block models. The
authors of \citet{decelle2011inference} describe an algorithm that
estimates block structure of a degree-corrected block model in time
linear in the number of vertices. Their algorithm is based on a
powerful heuristic of belief propagation from statistical physics. See,
for example, \citet{mezard2009information} for a survey level
treatment of belief propagation and a variety of applications. In the
context of sparse stochastic block models, these techniques have been
shown to be near optimal in estimating the underlying communities
[\citet{krzakala2013spectral}], at least in the balanced regime where
both communities are of equal size. A sublinear algorithm based on the
pseudo-likelihood of the sparse block model is described in \citet
{amini2013pseudo} wherein block labels are shown to be consistent in
the size of network. Finally, recent nonparametric representations of
the block model through dense graph limits, or graphons [\citet
{airoldi2013stochastic}] and network histograms [\citet
{olhede2013network}] provide promising new directions for the
understanding and estimation of block models.}

Another subclass of community detection methods are the so-called
extraction techniques where communities are extracted one at a time
[\citet{zhao2011community,lancichinetti2011finding}]. Rather than
search for an optimal partition or cover, these extraction methods seek
the strongest connected community sequentially. Extraction methods do
not force all vertices to be placed in a community and thereby are
flexible to loosely connected background vertices. ESSC is an
extraction method that utilizes the reference distribution of the
connectivity of a community based on the conditional configuration model.

{There are two main approaches currently used to assess the statistical
significance of communities in networks. The first approach, like ESSC,
builds upon statistical principles based on features of the observed
network itself. The second approach is permutation based in that the
significance of community structure is determined based on the results
of a prescribed method on many bootstrapped samples of the observed
network [see, e.g., \citet{clauset2008hierarchical}, \citet
{rosvall2010mapping}]. Many theoretical questions remain open for these
types of methods, including convergence of bootstrapped samples of networks.}
\subsection{Organization of the paper}
The remainder of this paper is organized as follows.
Section~\ref{sec:algorithm} is devoted to a detailed description of our
proposed algorithm for extraction
of statistically significant communities (ESSC), including motivation
and a description of the
reference distribution generated from the configuration model. In
Section~\ref{sec:relatedwork} we discuss the competing methods that we
use to validate our algorithm in both numerical and real network studies.
In Section~\ref{sec:realdata} we apply the ESSC algorithm to four
real-world networks. These results provide solid evidence that
ESSC performs well in practice, is competitive with (and in some cases
arguably superior to) several leading community detection
methods, and is effective in capturing background vertices.
In Section~\ref{sec:simulations}
we propose a test bed of benchmark networks for assessing the
performance of detection methods specifically on networks with
background vertices. To the best of our knowledge, this is the first
set of benchmarks proposed for networks of this type. We show that ESSC
outperforms existing methods on these background benchmarks. We also
show that ESSC performs competitively on standard (nonbackground)
benchmark networks with both nonoverlapping and overlapping community
structures. We end with a discussion of our work and avenues for future
research.

\section{The ESSC algorithm}\label{sec:algorithm}

\subsection{Conditional configuration model}

Let $G_o$ be an observed, undirected network having $n$ vertices.
Though many networks of interest will be
simple, $G_o$ may contain self-loops or multiple edges.
Assume without loss of generality that $G_o$ has vertex set $V = [n] =
\{1,2,\ldots,n\}$. The edge {multiset} $E_o$ of $G_o$
contains all (unordered) pairs $\{i,j\}$ such that $i,j \in[n]$ and
there is a link between vertices $i$ and $j$ in $G_o$,
with repetitions for multiple edges.
Let $d_o(u)$ denote the degree of a vertex $u$, that is, the number of
edges incident on $u$,
and let $\mathbf{d}_o = \{ d_o(1), \ldots, d_o(n) \}$ denote the degree
sequence of $G_o$.

The starting point for our analysis is a stochastic network model that
is derived from the degree
sequence $\mathbf{d}_o$ of $G_o$, specifically, the configuration
model associated
with $\mathbf{d}_o$, which we denote by $\operatorname{CM}(\mathbf{d}_o)$
[\citet{bender1978asymptotic}, \citet{bollobas1979probabilistic}, \citet
{molloy1995critical}].
The configuration model $\operatorname{CM}(\mathbf{d}_o)$
is a probability measure on the family of {multigraphs} with
vertex set $[n]$ and degree sequence
$\mathbf{d}_o$ that reflects, within the constraints of the degree
sequence, a random assignment of edges
between vertices.

The configuration model $\operatorname{CM}(\mathbf{d}_o)$ has a simple generative
form. Initially,
each vertex $u \in[n]$ is assigned $d_o(u)$ ``stubs,'' which act as
half-edges. At the next stage, two stubs are
chosen uniformly at random and connected to form an edge; this
procedure is repeated independently until
all stubs have been connected. Let $\widehat{G} = ([n], \hat{E})$
denote the random network generated by this
procedure.
Note that $\widehat{G}$ may contain self loops and multiple edges
between vertices,
even if the given network $G$ is simple.

The configuration model $\operatorname{CM}(\mathbf{d}_o)$ is capable of capturing and
preserving
strongly heterogeneous degree distributions often
encountered in real network data sets. {Importantly, all edge
probabilities in the configuration null model are determined solely by
the degree sequence $\mathbf{d}_o$ of an observed graph. As a result,
fitting a configuration model does not rely on simulation, rather,
estimation only requires the degree sequence of a single observed graph.}

Under the configuration model $\operatorname{CM}(\mathbf{d}_o)$ there are
no preferential connections between vertices, beyond what is dictated
by their degrees.
As such, $\operatorname{CM}(\mathbf{d}_o)$ provides a reference measure against
which we may assess the statistical significance of the connections
between two sets of vertices in the
observed network $G_o$: the more the observed number of cross-edges
deviates from the expected
number under the model, the greater the significance of the connection
between the vertex sets.
Let the observed network $G_o$ and the random network $\widehat{G}$ be
as above. Given a vertex $u \in[n]$
and vertex set $B \subseteq[n]$, {let}
\[
d_o(u: B) = \sum_{v \in B} \sum
_{e \in E_o} \mathbb{I}\bigl( e = \{ u,v\} \bigr)
\]
{denote} the number of edges between $u$ and some vertex in $B$
in $G_o$. Define $\hat{d}(u: B)$ {as} the
corresponding number of edges in $\widehat{G}$. Note that $\hat{d}(u
: B)$
is a random variable taking values in the set
$\{0, 1, \ldots, d_o(u) \}$, and that $d_o(u: B) = \hat{d}(u: B) =
d_o(u)$ when $B = [n]$ is the full vertex set.
We now state a theorem describing asymptotics for the random variable
$\hat{d}(u:B)$ in the configuration model which will form the basis of
the algorithm. Recall that the total variation distance between two
probability mass functions
$\vp:= \{p(i)\}_{i \geq0}$ and $\vq:= \{q(i)\}_{i \geq0}$ on
the space of natural numbers $\bN$
is defined~by
\[
d_{\mathrm{TV}}(\vp, \vq):= \frac{1}{2} \sum
_{i=0}^{\infty}\bigl |p(i) - q(i)\bigr|.
\]

%
\begin{theorem}
\label{prop:binomial}
{Let $\{\mathbf{d}_{o,n}\}_{n \geq1}$ be the degree sequences of an
observed sequence of graphs $\{G^n_o\}_{n \geq1}$, where $G^n_o$ is a
graph\vspace*{1pt} with vertex set $[n]$ and edge set $E_{o,n}$. Let $\{\widehat
{G}^n\}_{n\geq1}$ be the corresponding random graphs on $[n]$
constructed via the configuration model. Let $F_n$ be the empirical
distribution of $\mathbf{d}_{o,n}$.} Assume that there exists a cumulative
distribution function $F$ on $[0,\infty)$ with $0 < \mu:= \int_{\bR^+}
x\, dF(x) < \infty$
such that
%
%
\begin{equation}
F_{n} \weakc F
\end{equation}
and
%
%
\begin{equation}
\label{eq:A} \int_{\bR^+} x \,dF_{n}(x) \rightarrow
\mu.
\end{equation}
Fix $k \geq1$. For each $n \geq1$, let $u = u_n \in[n]$ be a vertex
with degree $d_{o,n}(u) = k$ and
let $B = B(n) \subseteq[n]$ be a set of vertices. Then the random
variable $\hat{d}_n(u: B)$ is approximately
$\operatorname{Binomial}(k, p_n(B))$ in the sense that
\[
d_{\mathrm{TV}}\bigl(\hat{d}_n(u: B), \operatorname{Bin}\bigl(k,
p_n(B)\bigr)\bigr) \to0,
\]
as $n\to\infty$. Here
%
%
\begin{equation}
\label{eq:setprobs} p_n(B) = \frac{ \sum_{v \in B} d_{o,n}(v)}{ \sum_{w \in
[n]} d_{o,n}(w)} = \frac{1}{2 |E_{o,n}|} \sum
_{v \in B} d_{o,n}(v),
\end{equation}
where $|E_{o,n}|$ is the total number of edges in the graph.
\end{theorem}
A precise proof of this fact is given in the Appendix~\ref{sec:theory}.
In light of the fact that the configuration model
$\operatorname{CM}(\mathbf{d}_o)$ does not contain preferential connections between
vertices, the probabilities
%
%
\begin{equation}
\label{eq:pval} p(u: B) = P\bigl( \hat{d}(u: B) \geq d_o(u: B)
\bigr)
\end{equation}
can be used to assess the strength of connection between a
vertex $u$ and a set of vertices $B \subseteq[n]$.
In particular, small values of $p(u: B)$ indicate that there are more
edges between $u$ and $B$ than
expected under the configuration model.

If we regard $d_o(u: B)$ as the observed value of a test statistic
that is distributed as
$\hat{d}(u: B)$ under the null model $\operatorname{CM}(\mathbf{d}_o)$, then $p(u: B)$
has the form of
a $p$-value for testing the hypothesis that $u$ is not strongly
associated with $B$.

This testing interpretation of $p(u: B)$ plays a role in the
iterative search procedure that underlies the ESSC method (see below).
However, we note that the testing point of view is informal, as the
null model $\operatorname{CM}(\mathbf{d}_o)$ itself
depends on the observed network $G_o$ through its degree distribution.

In general, the exact value of the probability $p(u: B)$ in (\ref
{eq:pval}) may be difficult to obtain.
In practice, the ESSC procedure approximates $p(u: B)$ by
$P( X_B \geq d_o(u: B) )$, where $X_B$ has a $\operatorname{Binomial}(d(u), p(B))$
distribution appealing to the result of Theorem~\ref{prop:binomial}.


\subsection{Description of the ESSC algorithm}

The core of the ESSC algorithm is an iterative {deterministic} procedure
(\emph{Community-Search}) that searches for robust, statistically
significant communities.
Beginning with an initial set $B_0$ of vertices that acts as a seed,
the procedure
successively refines and updates $B_0$
using (the binomial approximation of) the probabilities \eqref{eq:pval}
until it reaches a fixed point,
that is, a vertex set that is unchanged under updating.
The final vertex set identified by the search procedure is a detected community.

The \emph{Community-Search} procedure is applied repeatedly, using an
adaptively chosen sequence of seed vertices,
until it returns an empty community with no nodes. The resulting
collection ${\Cal C}$ of
detected communities (omitting repetitions) constitutes the output of
the algorithm. The seed set $B_0$ for the
initial run of the search procedure is the vertex of highest degree and
all of the vertices adjacent to it.
In subsequent runs of the search procedure the seed set $B_0$ is the
vertex of highest degree not contained
in any previously detected community and all the vertices adjacent to
it, regardless of whether the latter lie
in a previously detected community or not.

To simplify what follows, let $C_1,\ldots, C_K$ be the distinct
detected communities of $G_o$ in ${\Cal C}$.
The background of $G_o$ is defined to be
the set of vertices that do not belong to any detected communities:
%
%
\begin{equation}
\label{eq:def-bkgrnd} C_* = \operatorname{Background}(G_o \dvtx{\Cal C}) =
[n] {}\Big\backslash{} \bigcup_{k=1}^K
C_k.
\end{equation}
In principle, the number $K$ of detected communities can range from
zero to $n$.
Importantly, $K$ is not fixed in advance, but is adaptively determined
by the ESSC algorithm.
The identification of detected communities by the
\emph{Community-Search} procedure allows communities to overlap.
As with the number of discovered communities, $K$, the presence and
extent of overlap is
automatic; no prior specification of overlap specific parameters are required.

The updates of the \emph{Community-Search} procedure bear further
discussion. Consider an ideal setting in which,
for each vertex $u$ and vertex set $B$ we can determine, in an
unambiguous way, whether or not $u$
is strongly connected to $B$ in $G_o$.
Informally, a set of vertices $B$ is a community if the vertices $u \in
B$ have a strong connection
with vertices in $B$, while the vertices $u \in B^c$ do not.
Equivalently, $B$ is a community if and only if
it is a fixed point of the update rule
\[
S(A) = \bigl\{ \mbox{$u \in[n]$ such that $u$ is strongly connected with $A$}
\bigr\}
\]
that identifies the vertices having a strong connection with a set of
vertices $A \subseteq[n]$. Formally, we may
regard $S(\cdot)$ as a map from the power set of $[n]$ to itself. A vertex
set $B$ is a fixed point of $S(\cdot)$ if $S(B) = B$.
In order to find a fixed point of the update rule $S(\cdot)$, we apply the
rule repeatedly, starting from a seed set of vertices
$B_0$, until a fixed point is obtained. The eventual termination (and
success) of this simple procedure
is assured, as the power set of $[n]$ is finite. By the exhaustive or
selective considering of appropriate seed sets we
can effectively explore the space of fixed points of $S(\cdot)$, and thereby
identify communities
in $G_o$.

{The choice of a seed set $B_o$ for the \emph{Community-Search}
procedure requires further discussion. As currently implemented, we
choose $B_o$ as the neighborhood of the highest degree vertex among the
vertices lying outside currently extracted communities. Consider the
following situation, as pointed out by a referee, where one has two
disconnected clusters $C, C^\prime$ such that $C$ contains no inherent
community structure, for example, an \erdos random graph, and
$C^\prime $ contains strong community structure, for example, a
well-differentiated stochastic block model. If the maximal degree of
$C$ is larger than $C^\prime$, then ESSC could fail to find the
community structure in $C^\prime$. To address the above situation, one
can run the \emph{Community-Search} procedure in parallel across all
vertex neighborhoods. In this case, the final communities are the
collection of uniquely extracted vertex sets. We found that the
situation above did not arise in any of the applications or simulations
that we investigate in this paper.}

In practice, we make use of the probabilities $\{ p(u: B) \dvtx u \in[n]
\}$ to measure the
strength of the connection between $u \in[n]$ and $B$ relative to the
reference distribution $\operatorname{CM}(\mathbf{d})$.
In particular, we regard $p(u:B)$ informally as a $p$-value for testing
the null hypothesis
$\mathrm{H}^B_u$ that $u$ is not preferentially connected to $B$.
Then the task of identifying the vertices $u$ preferentially connected
to $B$ amounts to rejecting a subset of the
hypotheses $\{ \mathrm{H}^B_u \dvtx u \in[n] \}$. This is
accomplished in
steps 4 and 5 of the
\emph{Community-Search} procedure, where we make use of an adaptive
method of
Benjamini and Hochberg [\citet{benjamini1995controlling}] to reject a
subset of the hypotheses. The
rejection method ensures that the expected number of falsely rejected
hypotheses divided by the
total number of rejected hypotheses
(the so-called false discovery rate) is at most $\alpha$ [see \citet
{benjamini1995controlling} for more details].
A default false discovery rate threshold $\alpha$ of 5\% is common in
many applications, and we adopt this value
here. Pseudo-code for the \emph{Community-Search} procedure and ESSC
algorithm is shown below.

%
\begin{framed}
\begin{center}\emph{Community-Search Procedure}
\end{center}
\begin{itemize}
\item[] \textit{Given}: Graph $G_o = ([n], E_o)$; significance level
$\alpha\in(0,1)$.

\item[] \textit{Input}: Seed set $B_0 \subseteq[n]$.

\item[] \textit{Initialize}: $t:= -1$, $B_{-1} = \varnothing$.

\item[] \textit{Loop (Update)}: Until $B_{t+1} = B_t$
\begin{itemize}[\qquad]
\item[1.] $t:= t+1$.

\item[2.] Compute $p(u: B_t)$ for each $u \in[n]$.

\item[3.] Order the $n$ vertices of $G_o$ so that
$p(u_1: B_t) \leq\cdots\leq p(u_n: B_t)$.

\item[4.] Let $k \geq0$ be the largest integer such that $p(u_k: B)
\leq(k / n) \alpha$.

\item[5.] Update $B_{t+1}:= \{ u_1,\ldots, u_k \}$.
\end{itemize}

\item[] \textit{Return}: Fixed point community $B_t$.
\end{itemize}
\end{framed}

\begin{framed}
\begin{center}\emph{ESSC Algorithm}
\end{center}
\begin{itemize}
\item[] \textit{Input}: Graph $G_o = ([n], E_o)$; significance level
$\alpha\in(0,1)$.

\item[] \textit{Initialize}: $V = [n]$, ${\Cal C}:= \varnothing$.

\item[] \textit{Loop}:

\begin{itemize}
\item[] Let $u \in V$ be the smallest (in case of ties) vertex with
maximal degree.

\item[] Define seed set $B_0: = \{ u \} \cup\{ v \in[n] \dvtx\{
u,v \}
\in E_o \}$.

\item[] Obtain detected community $C:= \mbox{\emph
{Community-Search}}(B_0)$ from search procedure.

\item[] If $C \neq\varnothing$ then

\begin{itemize}
\item[] Update ${\Cal C}:= {\Cal C} \cup\{ C \}$.

\item[] Update $V:= V \setminus C$.

\item[] Repeat Loop.
\end{itemize}

\item[] Otherwise (if $C = \varnothing$), terminate the procedure.
\end{itemize}

\item[] \textit{Return}: Family ${\Cal C}$ of detected communities.
\end{itemize}
\end{framed}

\section{Competing methods}\label{sec:competingmethods}

Here we describe the set of community detection methods that we use for
validation and comparison with ESSC. We implement a variety of
{established} detection methods all of which have publicly available
code. {We note that we do not compare ESSC with the recently developed
fast block model algorithms from \citet {decelle2011inference}, \citet
{airoldi2013stochastic} and \citet {krzakala2013spectral}; such
comparisons would be interesting for future work.} The parameter
settings for each algorithm are described
in the Appendix~\ref{sec:parameters}.

\begin{longlist}[\quad]
\item[\textit{GenLouvain}:]  The GenLouvain method of \citeauthor{Jutla2011}\break (\citeyear
{Jutla2011}) is
a modularity-based method that employs an agglomerative optimization
algorithm to search for the partition that maximizes the score in (\ref
{eq:mod}). The algorithm is composed of two stages that are repeated
iteratively until a local optimum is reached. In the first, each vertex
is assigned to its own distinct community. Then for each vertex $u$ (of
community $B_u$), the neighbors of $u$ are sequentially added to $B_u$
if the addition results in a positive change in modularity. This
procedure is repeated for all vertices in the network until no positive
change in modularity is possible. In the second stage of the algorithm,
the communities found in the first stage are treated as the new vertex
set and passed back to the first stage of the algorithm where two
communities are treated as neighboring if they share at least one edge
between them. Throughout the remainder of this paper, we specify
$\widehat{G}_{\mathrm{null}}$ as the configuration model so that GenLouvain is
set to optimize the Newman--Girvan modularity [\citet
{newman2004finding}]. As a result, the Louvain methods of \citet
{blondel2008fast} and GenLouvain can be used interchangeably (notably,
however, the GenLouvain code does not exploit all possible efficiencies
for this null model).

\item[\textit{Infomap}:] The Infomap method of \citet
{rosvall2008maps} is a
flow-based method that seeks the partition that optimally compresses
the information of a random walk through the network. In particular,
the optimal partition minimizes the quality function known as the Map
Equation [\citet{rosvall2009map}], which measures the description length
of the random walk. The method employs the same greedy search algorithm
as Louvain [\citet{blondel2008fast}], refining the results through
simulated annealing.
\item[\textit{Spectral}:] Given a prespecified integer $k$, the Spectral
method of \citet{ng2002spectral} seeks the partition that best separates
the $k$ smallest eigenvectors of the graph Laplacian. Specifically, the
$k$ smallest eigenvectors of the graph Laplacian are stacked to form
the $n \times k$ eigenvector matrix $X$ and $k$-means clustering is
applied to the normalized rows of $X$. Vertices are then assigned to
communities according to the results of $k$-means. {We note that there
are proposed heuristics for choosing $k$. For example, the algorithm in
\citet{krzakala2013spectral} does not require one to specify the
number of communities in advance and uses the number of real
eigenvalues outside a certain disk in the complex plane as a starting
estimate. Throughout the manuscript, however, we choose $k$ based on
characteristics of the data investigated.}

\item[\textit{ZLZ}:] The method of \citet{zhao2011community}, which we
informally call ZLZ, is an extraction method that searches for
communities one at a time based on a local graph-theoretic criterion.
In each extraction, ZLZ employs the Tabu search algorithm [\citet
{glover1989tabu}] to find the community $B$ that maximizes the
difference of within-community edge density and outer edge density:
%
%
\begin{equation}
\label{eq:ZLZ} |B|\bigl| B^c\bigr|\sum_{{i,j} \in[n]} \biggl(
\frac
{A_{i,j} \mathbb{I}({i \in B, j \in B})}{|B|^2} - \frac{A_{i,j}
\mathbb
{I}({i \in B, j \in B^c})}{|B\Vert B^c|} \biggr),
\end{equation}
where $|B|$ denotes the number of vertices in $B$ and $A_{i,j}$ is the
$i$,$j$th entry of the adjacency matrix associated with the observed
graph. Once a community is extracted, the vertices of the community are
removed from the network and the procedure is repeated until a
prespecified number of disjoint communities are found. By following a
similar technique described in \citet{bickel2009nonparametric}, the
authors show that under a degree-corrected block model, the estimated
labels resulting from maximizing (\ref{eq:ZLZ}) are consistent as the
size of the network tends to infinity [see \citet{zhao2012consistency}
for more details].


%
\item[\textit{OSLOM}:]  The OSLOM method [\citet{lancichinetti2011finding}]
is an inferential extraction method that compares the local
connectivity of a community with what is expected under the
configuration model. Given a fixed collection of vertices $B$, the
method first calculates the probability of all external vertices having
at least as many edges as it has shared with the collection. These
probabilities are then resampled from the observed distribution. The
order statistics of the resampled probabilities are used to decide
which vertices should be added to $B$; a vertex is added whenever the
cumulative distribution function of its order statistic falls below a
preset threshold $\alpha$. Vertices are iteratively added and taken
away from $B$ in a stepwise fashion according to the above procedure.
This extraction procedure is run across a random set of initializing
communities and the final set of communities are pruned based on a
pairwise comparison of overlap.
\end{longlist}

%
\begin{table}[t]
\caption{A summary of the detection methods we consider in our
simulation and
application study. From left to right, we list the type of community
structure that each method can handle and the parameters required as
input for each algorithm. Listed free parameters include the following:
$k$, the number of communities; $\alpha$, the significance level; $N$,
the number of iterations; and $\gamma$, a resolution parameter}\label{tab:competing}
\begin{tabular*}{\textwidth}{@{\extracolsep{\fill}}lccccccc@{}}
\hline
& \multicolumn{3}{c}{\textbf{Community structure}} & \multicolumn
{4}{c@{}}{\textbf{Free parameters}}\\[-6pt]
& \multicolumn{3}{c}{\hrulefill} & \multicolumn
{4}{c@{}}{\hrulefill}\\
{\textbf{Method}} &\textbf{Disjoint} & \textbf{Overlapping} &
\textbf{Background} & $\bolds{k}$ & $\bolds{\alpha}$ & $\bolds{N}$ &
$\bolds{\gamma}$ \\
\hline
ESSC & $\checkmark$ & $\checkmark$ & $\checkmark$ & &$\checkmark$&
&
\\
OSLOM & $\checkmark$ & $\checkmark$ & $\checkmark$ & & $\checkmark$ &
$\checkmark$  \\
ZLZ & $\checkmark$ & & $\checkmark$ & $\checkmark$ & & $\checkmark$
\\
GenLouvain & $\checkmark$ & & & & & & $\checkmark$ \\
Infomap & $\checkmark$ & & & & $\checkmark$ & $\checkmark$ & \\
Spectral &$\checkmark$ & & & $\checkmark$ & & $\checkmark$ & \\
\hline
\end{tabular*}
\end{table}

\noindent There are a few similarities between ESSC and these described competing
methods. For instance OSLOM and GenLouvain both specify the
configuration model as a reference network model to which candidate
communities are compared. Both ZLZ and OSLOM are extraction methods,
like ESSC, that do not require all vertices to belong to a community.
The ESSC method uses the parametric distribution that approximates
local connectivity of vertices and a candidate community. {Since the
configuration model can be estimated using only the observed graph, the
probabilities in (\ref{eq:setprobs}) have a closed form which can be
computed analytically. On the other hand, OSLOM relies upon a
bootstrapped sample of networks for determining the significance of a
community. Whereas both OSLOM and ESSC are based on inferential
statistical techniques, Infomap, Spectral, ZLZ and GenLouvain use
network summaries  directly.} Unlike several of these mentioned methods,
ESSC requires no specification of the number of communities and only
relies upon one parameter which guides the false discovery rate.
We summarize the features of ESSC and these competing methods in
Table~\ref{tab:competing}.

%
\begin{table}[b]
\caption{Summary statistics of the four networks that we analyze}
\label{table:dataset}
\begin{tabular*}{\textwidth}{@{\extracolsep{\fill}}ld{6.0}d{6.0}@{}}
\hline
\textbf{Network} & \multicolumn{1}{c}{{\textbf{Number of vertices}}} &
\multicolumn{1}{c@{}}{{\textbf{Number of edges}}} \\
\hline
Caltech & 762 & 16\mbox{,}651 \\
Political blog & 1222 & 16\mbox{,}714 \\
Personal Facebook & 561 & 8375\\
Enron email & 36\mbox{,}691 & 293\mbox{,}307 \\
\hline
\end{tabular*}
\end{table}

%

\section{Real network analysis study}\label{sec:realdata}
Existing community detection methods differ widely in their underlying
criteria, as well as the algorithms they use to identify communities
that satisfy
these criteria. As such, we assess the performance of ESSC by comparing
it with several existing methods---OSLOM, ZLZ, GenLouvain, Infomap,
Spectral and $k$-means---on both a collection of real-world networks as well
as an extensive collection of simulation benchmarks.

\begin{figure}[t]

\includegraphics{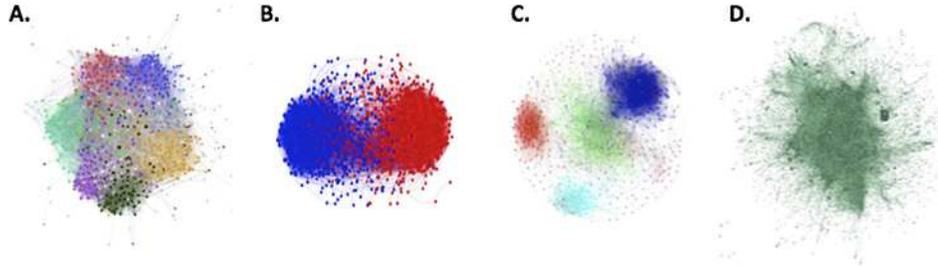}

\caption{Real networks analyzed in the paper. \textup{(A)} The Caltech
Facebook network of 2005 colored by dormitory residence. \textup{(B)} The
2005 political blog network colored by political affiliation. \textup{(C)}
The personal Facebook network of the first author colored by location
in which he met each individual. \textup{(D)} The Enron email network. Each
graph is drawn with the Force Atlas 2 layout using Gephi software.}
\label{fig:example}
\end{figure}

We first applied ESSC to four real networks of various size and
density: the Caltech Facebook network [\citet{traud2011comparing}], the
political blog network [\citet{adamic2005political}], the personal
Facebook network of the first author and the Enron email network [\citet
{leskovecENRON}]. We summarize the network structures in Table~\ref
{table:dataset} and visualize them in Figure~\ref{fig:example}.

%

On the first two networks, we compare quantitative features of the
communities of each method, including size, number of {communities},
extent of overlap and extent of background. Moreover, we
evaluate the ability of each method to capture specific features of
these two complex networks through a formal classification study. We
describe the precise settings of all tuning parameters for each of the
detection algorithms in the Appendix~\ref{sec:parameters}. All methods were
run on a 4 GB RAM, 2.8~GHz dual processor personal computer.

\subsection{Caltech Facebook network}\label{sec:caltech}

The Caltech Facebook network of \citet{traud2011comparing} represents
the friendship relations of a group of undergraduate students at the
California Institute of Technology on a single day in September, 2005.
An edge is present between two individuals if they are friends on Facebook.
In addition to friendship relations, several demographic features are
available for each student, including dormitory residence,
college major, year of entry, high school and gender. A summary of
these features is given in Table~\ref{table:features}.
This data set provides a natural benchmark for community detection methods
due to the possible association of community structure with one or more
demographic features.
Previous studies have found that this network displays community
structure closely matching the dormitory residence of the individuals
[\citet{traud2011comparing}]. We illustrate the network according to
residence in Figure~\ref{fig:example}({A}).

\subsubsection{Quantitative comparison} We first compare the
communities detected by each method based on quantitative summaries of
the communities themselves: the number and size of the communities; the
overlap present; and the number of background vertices found.
A summary of the findings is given in Table~\ref{table:caltechstats}.
ESSC took 1.584 seconds to run on this network.

%
%
\begin{table}[t]
\caption{A summary of the features associated with the individuals in
the Caltech Facebook network. From left to right, $k$ is the number of
unique categories, $p_{m}$ is the proportion of missing data, $m$ is
the minimum size of any unique category, and $M$ is the maximum size of
any unique category}
\label{table:features}
\begin{tabular*}{\textwidth}{@{\extracolsep{4in minus 4in}}ld{3.0}cd{3.0}
c@{}}
\hline
\textbf{Feature} & \multicolumn{1}{c}{$\bolds{k}$} &
\multicolumn{1}{c}{$\bolds{p_{m}}$}&
\multicolumn{1}{c}{$\bolds{m}$} &
\multicolumn{1}{c@{}}{$\bolds{M}$}\\
\hline
Dormitory & 8 & 0.2205 & 44 & \phantom{0}98 \\
Year & 15 & 0.1457 & 1 & 173 \\
Major & 30 & 0.0984 & 1 & \phantom{0}88 \\
High school & 498 & 0.1693 & 1 & \phantom{00}3 \\
Gender & 2 & 0.0827 & 227 & 472 \\
\hline
\end{tabular*}\vspace*{-3pt}
\end{table}
%

%
%
\begin{table}[b]
\caption{A summary of the detection methods run on the Caltech Facebook
network. From left to right,
$N_C$ is the number of communities detected,
$\overline{S}$ is the average size of the communities,
$\hat{\sigma}_S$ is the standard deviation of the community size,
$\overline{\mathbb{M}}$ is the average number of communities to which
nonbackground vertices belong,
$\overline{D}_{\mathrm{sig}}$ is the average degree of the vertices in a community,
$\overline{D}_{B}$ is the average degree of the background vertices,
$P_B$ is the proportion of background vertices, and $\hat{E}$ is the
mean classification error associated with the dormitory feature of the
individuals.
*Methods were set to find 7 and 8 communities, based on the number of
communities
detected by ESSC and GenLouvain. ---: represents repeated values}
\label{table:caltechstats}
\begin{tabular*}{\textwidth}{@{\extracolsep{\fill}}ld{2.0}d{3.2}d{3.2}ccccc@{}}
\hline
\textbf{Method} & \multicolumn{1}{c}{$\bolds{N_C}$} &
\multicolumn{1}{c}{$\bolds{\overline{S}}$} &
\multicolumn{1}{c}{$\bolds{\hat{\sigma}_S}$} &
\multicolumn{1}{c}{$\bolds{\overline{\mathbb{M}}}$} &
\multicolumn{1}{c}{$\bolds{\overline{D}_{\mathrm{sig}}}$} &
\multicolumn{1}{c}{$\bolds{\overline{D}_{B}}$} & \multicolumn{1}{c}{$\bolds{P_B}$} &
\multicolumn{1}{c@{}}{$\bolds{\hat{E}}$} \\
\hline
ESSC & 7 & 78.57 & 16.03 & 1.034 & 55.75 & 15.81 & 0.3018 &
0.0925\\
OSLOM & 18 & 86.78 &63.25 & 1.085 & 50.30 & \phantom{0}6.18 & 0.1496 & 0.2011\\
ZLZ* & 7 & 62.14 &41.97 &1\phantom{000.} &64.08 & 16.60 & 0.4291 & 0.5346 \\
ZLZ* & 8 & 58 &40.58 & -- & 62.44 & 14.53 & 0.3911 & 0.5323 \\
GenLouvain & 8 & 95.25 & 35.75 & -- & 43.70 & NA & NA & 0.2576 \\
Infomap & 18 & 42.33 & 46.23 &-- & -- & -- & -- & 0.8132\\
Spectral* & 7 & 108.86& 72.77 & --& -- & -- & --& 0.4865 \\
Spectral* & 8 & 95.25 & 61.52 &-- & -- & -- & -- & 0.4512\\
$k$-means* & 7 & 108.86& 126.51 &-- & -- & -- & --& 0.4242\\
$k$-means* & 8 & 95.25 & 118.35&-- & -- & -- & -- &0.4327\\
\hline
\end{tabular*}
\end{table}

%
%
\begin{figure}

\includegraphics{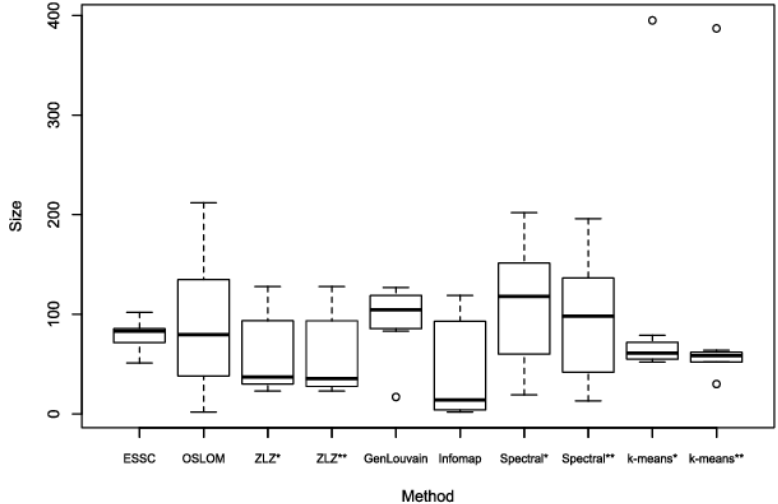}

\caption{The size distributions of communities from each detection
method when run on the Caltech network.}\label{fig:caltechsize}
\end{figure}

%
We note that the ZLZ, $k$-means and Spectral methods require prior
specification of the number of discovered communities.
Based on the ESSC and GenLouvain
results, we ran each of these methods with seven and eight detected
communities. We show the size distributions of the detected communities
for each method in Figure~\ref{fig:caltechsize}, and find that the size
distribution is broadly similar across the ESSC, ZLZ, GenLouvain and
Spectral methods. Infomap found many ($N_C=18$) small communities,
including several communities of size three or fewer. At both $k = 7$
and 8, $k$-means found one large community as well as many small
similarly sized communities. Interestingly, GenLouvain also produced an
eighth community of size twenty-one, all of whose vertices were part
of the background vertex set determined by ESSC.
No method found significant overlap among the detected communities.
The average number of communities to which each vertex belonged ranged
from 1 to 1.085.
Each of the methods capable of detecting background (ESSC, OSLOM and
ZLZ) designated more than
$15 \%$ of the total network as background, and vertices contained
within communities had average degree
nearly three times that of background vertices. This suggests, as
expected, that the background vertices are
less connected to other vertices in the network.


\subsubsection{Community features}\label{sec:adaboost}
One motivation for community detection\break methods is their ability to
find communities of vertices that represent interesting, but possibly
unavailable, features of the system under study.
Here, we explore the ability of each method to capture the demographic
features of the Caltech network.
To do this, we measure the extent to which the demographic features
``cluster'' within communities.
Typical pair counting measures do not work well here, as the detected
communities may overlap and may not
cover the entire network. Also, pair counting measures treat the
features as a ``ground truth'' partition of the network,
whereas the true structure of a network is often more complex [\citet
{Yang2012Groundtruth,lee2013benchmarking}].
As an alternative, we address the connection between communities and
features through the problem of
classification [see, e.g., \citet
{shabalin2009finding,friedman2001elements}]: for each vertex, we treat its
community identification as a predictor and its demographic features as
a discrete response that we wish to predict.
We describe our approach in more detail.

Suppose that a detection method divides the vertices of the network
into $K$ communities plus background.
Then the $n \times K$ matrix $X = [x_{i,j}]$ defined~by
\[
x_{i,j} = %
\cases{ 1, &\quad $\mbox{if vertex $i$ belongs to
community $j$},$\vspace*{2pt}
\cr
0, &\quad $\mbox{otherwise},$} %
\]
represents the detected community structure of the network.
For a given demographic feature $\alpha$ taking $L$-values, let
$y^{\alpha}_i \in[L]$ be the value of $\alpha$
in sample $i$. We ignore samples for which the value of feature $\alpha
$ is not available.
Treating the $i$th row of the matrix $X$ as a $K$-variate predictor for
$y^\alpha_i$, we use the Adaboost classification method [\citet{adaboost1997}]
with tree classifiers to construct a prediction rule $\phi\dvtx \{
0,1\}^K
\to[L]$.

%
%
\begin{figure}[b]

\includegraphics{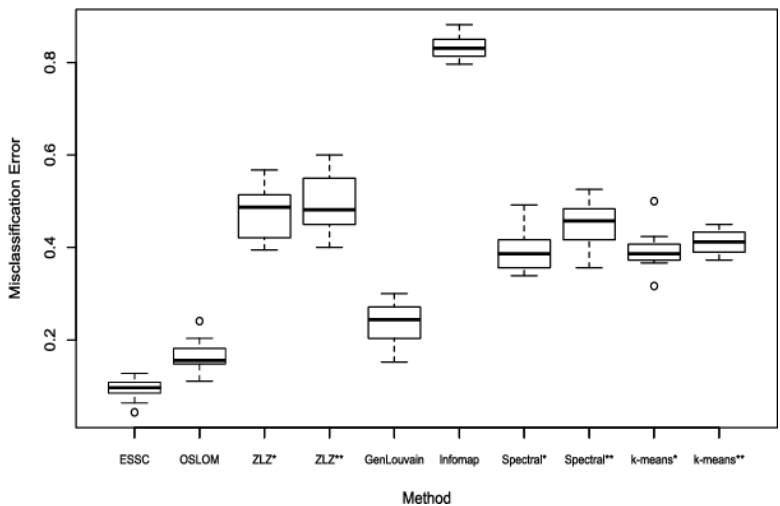}

\caption{The misclassification error of each method based on the
ten-fold classification study performed on the Caltech network. The
community containment of each individual was used to classify his/her
dormitory residence. For each test, an Adaboost classifier was used for
comparison.}\label{fig:caltecherror}
\end{figure}

To evaluate each method, we first randomly divide the $n$ samples into
ten equally sized subgroups. Then by setting aside one subgroup as a
test set, we train the classifier on the remaining subgroups and
predict the features of the test set. By subsequently treating each
subgroup as a test set in this way, we calculate the misclassification
error associated with each test. We report the average
misclassification error $\hat{E}$ for each method as a means of
comparison and report the results in Table~\ref{table:caltechstats}.
The distribution of errors is shown in Figure~\ref{fig:caltecherror}.
Values of $\hat{E}$ near zero suggest that the detected\vadjust{\goodbreak} community
structure captures the clustering of the selected feature. We consider
the dormitory residence of the network, as this feature has been shown
to be most representative of the community structure in past studies
[\citet{traud2012social}].
From Figure~\ref{fig:caltecherror}, we see that ESSC has the lowest
misclassification error among competing methods in this classification
study. These results suggest that the detected communities of ESSC best
match the dormitory residence of the Caltech network.


%

\subsection{Political blog network}

The political blog network of \citet{adamic2005political} represents the
hyperlink structure of 1222 political blogs in 2005 near the time of
the 2004 U.S. election. Undirected edges connect two blogs that have at
least one hyperlink between them. The blogs were pre-classified
according to political affiliation by the authors in \citet
{adamic2005political}. These authors, as well as those of \citet
{newman2006modularity}, observed that blogs of a similar political
affiliation tend to link to one another much more often than to blogs
of the opposite affiliation. We show a force directed layout of this
network colored by political affiliation in Figure~\ref{fig:example}({B}).

%
%
\begin{table}[b]
\caption{A summary of the detection methods run on the Political blog
network. The statistics shown here are the same as those in Table
\protect\ref
{table:caltechstats}. *We set $k$ to 2 to match the results of
GenLouvain and ESSC. **We chose $k$ as 10 so that at least 50 percent
of the vertices were placed in a community}
\label{table:adamicstats}
\begin{tabular*}{\textwidth}{@{\extracolsep{4in minus 4in}}ld{2.0}d{3.2}d{3.2}cd{1.3}d{1.3}d{1.4}d{1.4}@{}}
\hline
\textbf{Method} & \multicolumn{1}{c}{$\bolds{N_C}$} &
\multicolumn{1}{c}{$\bolds{\overline{S}}$} &
\multicolumn{1}{c}{$\bolds{\hat{\sigma}_S}$} &
\multicolumn{1}{c}{$\bolds{\overline{\mathbb{M}}}$} &
\multicolumn{1}{c}{$\bolds{\overline{D}_{\mathrm{sig}}}$} &
\multicolumn{1}{c}{$\bolds{\overline{D}_{B}}$} &
 \multicolumn{1}{c}{$\bolds{P_B}$} &
\multicolumn{1}{c@{}}{$\bolds{\hat{E}}$} \\
\hline
ESSC & 2 & 448.50 & 75.66 & 1\phantom{000.} & 36.322 & 2.577 & 0.2651 & 0.0201 \\
OSLOM & 11 & 87.58 & 79.48 &1.110 & 33.749 & 5.342 & 0.225 & 0.0306\\
ZLZ** & 10 & 60.00 &37.69 &1\phantom{000.} &35.50 &2.50 &0.506 &0.1341 \\
GenLouvain & 2 & 611.00 & 72.12 &-- & 27.36 & \multicolumn{1}{c}{NA} & 0 & 0.0475\\
Infomap & 36 & 33.94 & 125.74& -- & \multicolumn{1}{c}{--} & \multicolumn{1}{c}{--} & \multicolumn{1}{c}{--} & 0.0532 \\
Spectral* & 2 & 611.00 & 858.43 & -- & \multicolumn{1}{c}{--} & \multicolumn{1}{c}{--} & \multicolumn{1}{c}{--} & 0.3821\\
$k$-means* &2 & 611.00 & 613.77 & -- & \multicolumn{1}{c}{--} & \multicolumn{1}{c}{--} & \multicolumn{1}{c}{--} & 0.2856 \\
\hline
\end{tabular*}
\end{table}


\subsubsection{Quantitative comparison} We first compare the
communities detected by each method based on their quantitative characteristics.
The results are summarized in Table~\ref{table:adamicstats}. ESSC took
2.012 seconds to run on this network.

Both the ESSC algorithm and GenLouvain found two large communities of
similar size.
Interestingly, Infomap found thirty-six communities, thirty-four of
which contained fewer than 25 vertices. Roughly $95\%$ of
the vertices in these smaller communities of Infomap were contained in
the background vertices of ESSC.
Neither ESSC nor OSLOM found significant overlap among the communities,
reflecting the tendency of the political bloggers
to communicate with like-minded individuals: as noted by the authors of
\citet{adamic2005political}, ``divided they blog.''

ESSC, OSLOM and ZLZ each assigned over twenty percent of the vertices
to background. The pairwise Jaccard score of
these background sets
is greater than 0.67 in each case.
The background vertices of all three extraction methods had mean degree
six times smaller
than vertices within communities, suggesting the presence of sparsely
connected background vertices in this network.

\subsubsection{Political affiliation} We now evaluate the extent to
which the political affiliation of the blogs ``cluster'' by conducting
the same classification study detailed in Section~\ref{sec:adaboost}.
We report the mean proportion of misclassified labels $\hat{E}$ in
Table~\ref{table:adamicstats}.
ESSC, OSLOM, GenLouvain and Infomap all maintained classification
errors below $10\%$,
suggesting that political affiliation is captured by the network's
community structure quite well.
ESSC had the lowest misclassification error in this study, keeping an
error below $4\%$ across all tests. We look deeper into the strength of
connection of the background vertices to the true political
affiliations. Interestingly, these vertices were still preferentially
attached to their true affiliation, however, their associated $p$-values
were typically greater than 0.10, indicating weak affiliation.

\subsection{Personal Facebook network}\label{sec4.3}
The personal Facebook network gives friendship structure of the first
author's friends on Facebook.
In addition, each individual is labeled according to the time period
during which {he or she} met the first author. This data set, as
well as the labels, is provided in the supplemental file [\citet{supplemental}].
This network is shown, colored by label, in Figure~\ref{fig:example}({C}).

The understanding of human social interactions has been improved
through the analysis of large available social networks like Facebook
[\citet{lee2013benchmarking,traud2011comparing,traud2012social}].
Typically, these networks capture the social activity of individuals of
a single location. For example, the Facebook network analyzed in
Section~\ref{sec:caltech} reflects the friendships of individuals
specifically from the California Institute of Technology. The personal
Facebook network provides one view of how individuals from different
schools and locations
interact given that they all have one friend in common.

We ran ESSC on the network (running time about 1 second) and found 7 communities
with sizes varying from 10 to 157; see Table~\ref{table:personalfb}.
Approximately $18\%$ of the nodes in the network were distinguished as
background.
The mean degree of~the vertices belonging to a community ($\overline
{D}_{\mathrm{sig}} \approx33$) was about seven times that of the background
($\overline{D}_{B} \approx5$). Of the vertices that were contained in
a community, the average membership was very close to 1, suggesting
little overlap between communities.

%
\begin{table}[t]
\caption{Features of the personal Facebook network as well as the
results of ESSC. On the left, we list the labels of the individuals
according to location and the size of each group. On the right, we list
the detected communities and background as well as their corresponding
size}\label{table:personalfb}
\begin{tabular*}{\textwidth}{@{\extracolsep{\fill}}ld{3.0}cd{3.0}@{}}
\hline
\multicolumn{2}{@{}c}{\textbf{True features}} &
\multicolumn{2}{c@{}}{\textbf{ESSC results}} \\[-6pt]
\multicolumn{2}{@{}c}{\hrulefill} &
\multicolumn{2}{c@{}}{\hrulefill} \\
\textbf{Label} & \textbf{Size} & \textbf{Community} & \multicolumn{1}{c@{}}{\textbf{Size}}
\\
\hline
Acquaintance & 80 & 1 &43 \\
A &62 & 2 & 107 \\
B & 94& 3 & 75\\
C & 150& 4 & 157\\
D & 147& 5 & 53 \\
E & 3& 6 & 26\\
F & 3& 7 & 10\\
G & 22 & Background & \multicolumn{1}{c@{}}{101 (18.0$\%$)}\\
\hline
\end{tabular*}
\end{table}

%
\begin{figure}[b]

\includegraphics{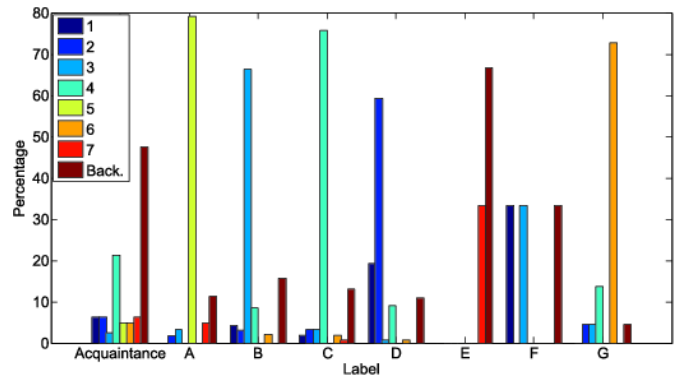}

\caption{A bar plot showing the clustering of locations A--G and
Acquaintances of the personal Facebook network. For each location
label, we show the percentage of individuals from that location that
were contained in each detected community. Communities are labeled 1--7
and Back. represents the background vertices.}\label{fig:personalfb}
\end{figure}

To understand how the location feature of the individuals cluster, we
investigate the composition of each label according to detected
community in Figure~\ref{fig:personalfb} and find several interesting results.
The individuals from locations A, B, C, D and G all tend to cluster
according to the detected communities. For instance, $79\%$ of the
individuals from location A were contained in community 5. Similarly,
60$\%$ or more of the individuals from locations B, C, D and G also
belong to a single community in each case. Groups A, B, C and D
represent the schools that the author attended from high school to
final graduate school and make up nearly 81$\%$ of the total network.
Groups E and F are not captured well by the communities, however, this
is expected due to the small size of these locations ($n = 3$ in both
cases). Finally, the most highly represented group among the background
distinguished by ESSC were
acquaintances---individuals met through other friends, events or
conferences. These results suggest that friendships in this network
cluster are based on location and that the acquaintances of the author
are not well connected to his remaining friends.

\subsection{Enron email network}
The Enron email network from \citet{leskovecENRON} is a large (36,691
vertices), sparse network in which
each vertex represents a unique email address.
An undirected edge connects any two addresses if at least one email
message has been sent from one address
to the other. At least one vertex of each edge corresponds to the email
address of an employee of the Enron corporation.
The network is shown in Figure~\ref{fig:example}(D). {We ran ESSC on
the network with $\alpha= 0.05$. ESSC took approximately 10 minutes to
run on this network.}

Importantly, the network includes Enron employees as well as
advertising agencies and spam sites outside Enron.
As such, we expect there to be many background vertices representing
spam and advertisement email addresses. On applying ESSC to the
network, we indeed find an abundance of background vertices---nearly
83$\%$ (30,454 vertices) of the network. The average degree of the
vertices within a community is nearly twelve times that of the
background vertices. ESSC found 8 communities with average size of 1239
and standard deviation 450.
The average membership of the vertices that were contained within a
community was 1.409, indicating a moderate amount of overlap of communities.

\section{Simulation study}\label{sec:simulations}
In this section we evaluate the performance of ESSC on simulated
networks with three primary types of community structure:
(1) communities that partition the network; (2) communities that
overlap and cover the network; and
(3) disjoint communities plus background.

Networks of the first two types have been well studied, and there are
several existing simulation benchmarks
for these structures [\citet{girvan2002community},
\citeauthor{lancichinetti2009benchmarks} (\citeyear{lancichinetti2009benchmarks,lancichinetti2009community})].
We make use of the Lancichinetti, Fortunato and Radicchi (LFR)
benchmark from\break \citeauthor{lancichinetti2009benchmarks} (\citeyear{lancichinetti2009benchmarks,lancichinetti2009community}) in order to
assess the performance of ESSC
and other methods on networks of the
first two types. Our principal reason for using the LFR simulation
benchmark is its flexibility, as well as the fact that
the power-law degree distribution it employs is representative of the
degree of heterogeneity present in many real networks [\citet
{barabasi1999emergence}]. ESSC performs well on these standard
nonoverlapping and overlapping benchmarks, and is in fact competitive
with the other detection methods in these settings. We evaluate the
results on these benchmarks in the Appendix~\ref{sec:nonbacksims}.

Relatively little attention has been paid to networks with background
vertices, and we are not aware of
a simulation benchmark for networks of this sort.
We therefore propose a flexible simulation benchmark for networks with
background that
extends the LFR benchmark, and use it to compare ESSC with competing methods.

%
%
\begin{table}[b]
\caption{Description of the free parameters available with the LFR
benchmark networks}\label{tab:LFR}
\begin{tabular*}{\textwidth}{@{\extracolsep{\fill}}ll@{}}
\hline
\textbf{Parameter}& \multicolumn{1}{c@{}}{\textbf{Description}}\\
\hline
$n$ & Size of the network\\
$\mu\in(0,1)$ & Mixing parameter: the proportion of external\\
& \quad community degree for each vertex\\
$\tau_1$ & Power-law exponent for degree distribution of network\\
$\tau_2$ & Power-law exponent for size distribution of communities in
network\\
$\bar{D}$ & Mean degree \\
$[s_1,s_2]$ & Size range of each community: $s_1 = {}$lower limit\\
& \quad$s_2 = {}$upper limit \\
$\rho\in(0,1)$ & Proportion of vertices contained in two communities
\\
& \quad(used in overlapping benchmark only)\\
\hline
\end{tabular*}
\end{table}

In the remainder of the section, we first describe the LFR benchmarks
of \citeauthor{lancichinetti2009benchmarks} (\citeyear{lancichinetti2009benchmarks,lancichinetti2009community}) and
then show how these benchmarks can be extended to networks with
background. We assess the performance of ESSC and other competing
methods on networks with background using our proposed benchmark.

\subsection{The LFR benchmark}\label{sec:LFR}
The LFR benchmarks of Lancichinetti and\break Fortunato
(\citeyear{lancichinetti2009benchmarks,lancichinetti2009community}) include a
number of parameters that govern
the community structure of the simulated network; a list is given in
Table~\ref{tab:LFR}.
The edge density of the simulated network is controlled through the
size $n$ of the network and the mean degree $\bar{D}$.
For example, sparse networks are represented by benchmarks with large
$n$ and small $\bar{D}$. The degree distribution
of simulated networks follows a power law with exponent $\tau_1$. Lower
and upper limits of the degree distribution are set to maintain an
average degree $\bar{D}$ among vertices in the network.
The distribution of community sizes in the LFR benchmark follows a
power law with exponent $\tau_2$. The size range $[s_1,s_2]$ sets lower
and upper limits on the size of communities in the network.
Consider a vertex $u$ and its community $C$. Then $u$ shares a fraction
$\mu$ of its edges with vertices outside of $C$ while the remaining
$1-\mu$ of its edges are shared with vertices within $C$. Thus, the
mixing parameter $\mu$ controls the extent to which communities mix,
with communities becoming less distinguishable as $\mu$ increases.
Finally, in the LFR benchmark with overlap,
the parameter $\rho\in(0,1)$ is the proportion of vertices that are
contained in exactly two communities, and
therefore controls the extent of overlap.
If $u$ belongs to two communities in the overlapping LFR benchmark,
then $\mu$ represents the
proportion of edges of $u$ that fall outside all these communities.

%

\subsection{Background benchmarks}

To assess detection methods on networks with background, we propose
three principled test bed simulations: (1) a network with no
communities (and therefore all vertices are background); (2) a network
with a single embedded community; and (3) a network with disjoint
communities and background. In what follows, we first describe how to
simulate each type of network and then discuss the results for each
type.

\textit{Networks with no community structure}: It is important to measure
the extent to which a detection method correctly identifies the lack of
community structure when none is present. We construct such background
networks by using two random network models:
the \erdos model of \citet{erdHos1960evolution} where all vertices are
linked with equal probability, and the configuration model of \citet
{molloy1995critical} where vertices are linked according to a
prescribed degree sequence as discussed in Section~\ref{sec:algorithm}.

For each of these models, we vary the size $n$ and mean degree $\bar
{D}$ in order to control the edge density of the generated network. In
particular, for configuration random networks, we specify that the
degree sequence follows a power law with degree $\tau_1$ and average
degree $\bar{D}$.

\textit{Single embedded community}: We consider networks that contain a
single embedded community and many background vertices.
To construct such networks, we use a variant of the stochastic two
block model of \citet{snijders1997estimation}, that has a simple
generative procedure. First, vertices are placed randomly and
independently in two blocks, $C_1$ and $C_2$, according to the
probabilities $\pi_1$ and $\pi_2 = 1- \pi_1$.
An edge is included between a pair of distinct vertices $u \in C_i$ and
$v \in C_j$ with probability $P_{i,j}$,
independently from pair to pair.

To construct a network of size $n$ with a single embedded community
$C_1$ and background $C_2$,
we generate a stochastic two block model
using $\bolds{\pi} = \{\pi, 1 - \pi\}$ with $\pi\in(0,1)$ and
$\textbf{P} = \{ P_{i,j} \dvtx1 \leq i,j \leq2 \}$ given by
\[
\mathbf{P} = \theta %
\pmatrix{ \kappa& 1
\cr
1 & 1 } %
.
\]
Here $\kappa> 1$ controls the inner community edge probability, and
$\theta< 1$
controls the average degree of the network.
Modifying $\pi$ controls for the size of the embedded community. The
parameters $\theta$ and $n$ can be modified to control the edge density
of the network. By generating a network of fixed size and mean degree,
one can assess the sensitivity of a detection method by running the
method across a range of $\pi$. We note that \citet{zhao2011community}
used a similar benchmark network to assess the performance of their own
detection algorithm.

\textit{Disjoint communities and background}: As a third benchmark test
set, we simulate a network with background and degree heterogeneities.
To do so, we propose combining the LFR benchmark described in
Section~\ref{sec:LFR} with the block structure described above. We construct
this network in two steps using the same parameters as the LFR
benchmark described in Table~\ref{tab:LFR}. First, we independently and
randomly assign vertices to one of two blocks $C_1$ and $C_2$ according
to probabilities $\bolds{\pi} = \{ \pi, 1- \pi\}$.
We place edges between vertices in block $C_1$ according to the
disjoint LFR benchmark with parameters
$\Theta= (\tau_1, \tau_2, n \cdot\pi, \mu, \bar{D} \cdot\pi,
[s_1,s_2])$. The remaining vertices, corresponding to
$C_2$, are connected to all vertices with equal probability $P_2:=
\bar
{D} (1 - \pi)$.
Thus, our benchmark is constructed as a stochastic 2 block model
described by $\bolds{\pi}$ and
\[
\mathbf{P} = %
\pmatrix{ \mathbf{P}_{\mathrm{LFR}} & P_2
\cr
P_2 & P_2 } %
,
\]
where $\mathbf{P}_{\mathrm{LFR}}$ denotes the edge probabilities between vertices
in $C_1$
derived from the LFR random network.
The resulting network has average degree $\bar{D}$. On average, a
fraction $\pi$ of the vertices
exhibit community structure following the LFR disjoint benchmark, while
the remaining vertices are
connected to each other and to vertices in the first block in an \erdos
like fashion.
This new benchmark is flexible and can be used to assess the
performance of any community detection
method for networks with background.

\subsection{Results}

\textit{Networks with no community structure}:
We generated both \erdos and configuration model random graphs with
1000 vertices, with average degree
$\bar{D}$ ranging from 10 to 100 in increments of 10. The degree
sequence of the vertices in the configuration network follow a
power-law distribution with degree $\tau_1 = 2$. For each value of
$\bar
{D}$, we generate 30 random graphs,
with edge probabilities determined by the value of $\bar{D}$.
In each of the simulations, ESSC assigned all nodes to background, as
desired.

\textit{Single embedded community}:
We generated networks of size 2000, and set $\kappa$ to~10, so that the
edge probability within the
single community is ten times that of the background.
We selected values of $\theta$ to generate networks with average degree
$\bar{D}$ of 30, 40 and 50.
For each value of $\bar{D}$, we generated networks with embedded
communities of size
$\pi* 2000$ for $\pi$ ranging from 0.01 to 0.3.

For each set of parameters, we generated 30 network realizations and
gave these as input to ESSC, Spectral, ZLZ and OSLOM. We set Spectral
to partition the network into two communities and set ZLZ to extract
one community, thereby giving both of the methods an advantage over the
other methods considered.

\begin{figure}

\includegraphics{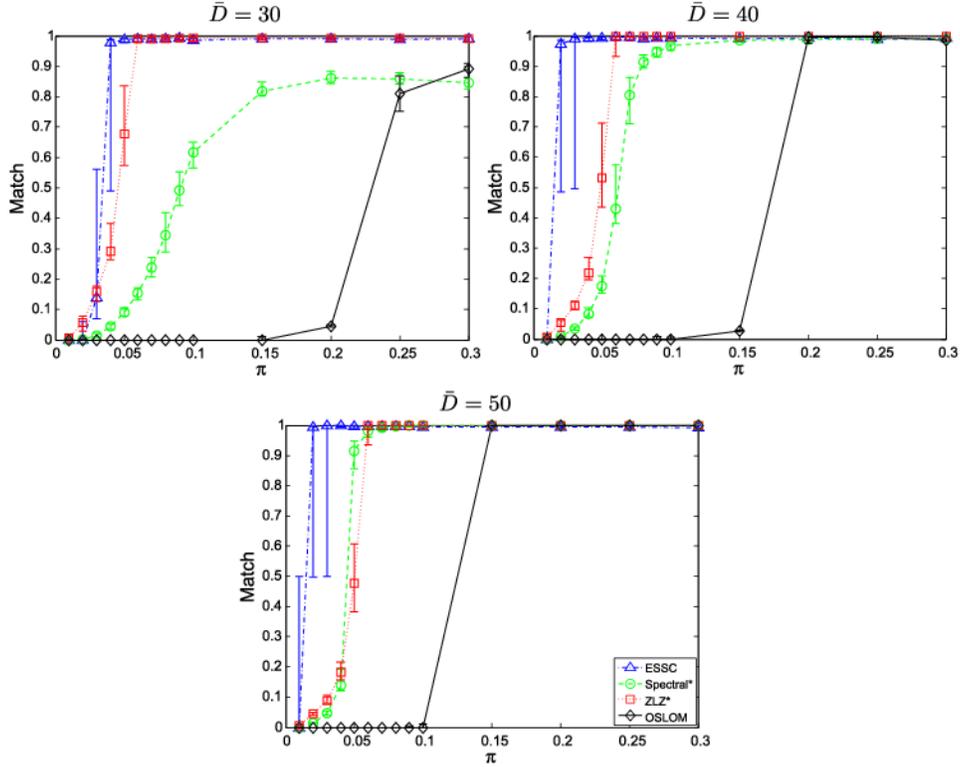}

\caption{The results for networks with a single embedded community.
Shown are the first, second and third quartile
of the maximum Jaccard Match of each method over 30 realizations across
values of $\pi$.
*Spectral and ZLZ were given the true number of communities: Spectral
was set to partition the
network into two communities, while ZLZ was set to extract 1
community.}\label{fig:Singleembed}
\end{figure}

In order to measure the ability of each method to find the true single
embedded community, we used the maximum Jaccard Match score of the
detected communities. In detail, we measured the Jaccard score between
each detected community and the true embedded community and reported
the maximum of these values for each simulation. Results are shown in
Figure~\ref{fig:Singleembed}.\looseness=-1

%
%

%
%

From Figure~\ref{fig:Singleembed}, we see that ESSC is able to find,
with Match $\approx$ 1, single embedded communities even when the
community is as small as $4\%$ of the total network. As the size of
embedded community increases, the performance of each method improves,
eventually reaching near optimal performance. In the case of small
embedded communities ($\pi< 0.05$), ESSC and ZLZ perform similarly,
with ESSC having a slight advantage. Finally, ESSC and all other
methods improve as the average degree of the network increases. Across
all simulations, we note that OSLOM did not find more than two
nontrivial communities.

\begin{figure}

\includegraphics{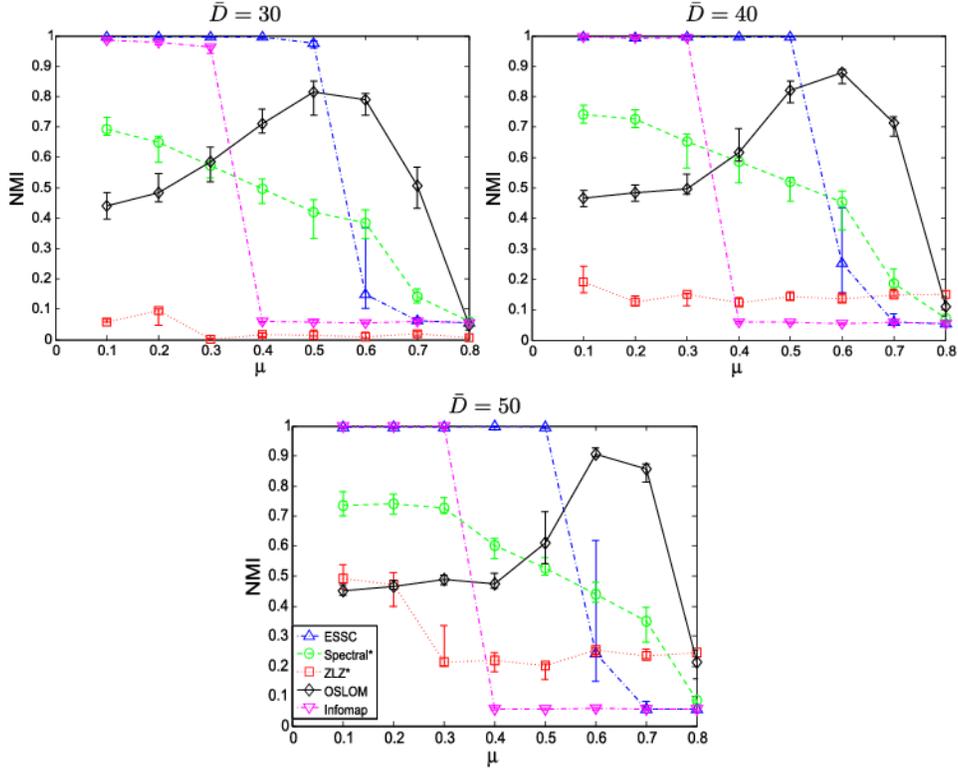}

\caption{The results for networks with LFR and background features.
Shown are the first, second and third quartile match of each method
over 30 realizations across values of $\mu$. The degree distribution of
the significant community structure follows a power law with exponent
$\tau_1 = 2$ with average degree $\bar{D}$ specified in each
figure. *Here, Spectral and ZLZ were given the true number of
communities.}\label{fig:LFRextraction} 
\end{figure}

\textit{Disjoint communities and background}: We simulated networks of
size $n = 2000$ with
$\pi= 1/2$, so that half of the vertices were background and the other
half belonged to disjoint communities
generated according to the LFR benchmark. Networks were generated with
average degree
$\bar{D} = 30$, 40 and 50, with community sizes in the range $[s_1,s_2]
= [20,100]$. Degree distributions were generated according to a power
law with degree exponent $\tau_1 = 2$ and community size distributions
were generated according to a power law with degree exponent $\tau_2 = 1$.
For each value of $\bar{D}$, networks were generated with mixing
parameter $\mu$ ranging between
0.1 and 0.8 in increments of 0.1.
For each set of parameters 30 network realizations were generated and
then passed as input to
ESSC, Spectral, ZLZ, OSLOM and Infomap. As before, the Spectral and ZLZ
were run using the true number
of communities. The generalized normalized mutual information (NMI) was
used to measure the concordance of the detected communities
and the true communities {with} background vertices treated as a
single community. NMI is an information theoretic tool that can measure
the similarity between two partitions as well as between two covers of
a network. For more information on this similarity measure, refer to
\citet{lancichinetti2009detecting}.
Results are shown in Figure~\ref{fig:LFRextraction}.

%
%

%
%
%

Figure~\ref{fig:LFRextraction} tells us several interesting things
about the performance of ESSC and other detection methods on complex
networks with background. First, we see that ESSC performs well (NMI
$\approx1$) across a range of mixing parameters $\mu$ from 0.1 to 0.5.
After $\mu= 0.6$, ESSC finds no significant communities and, hence,
the performance falls at this point. Infomap competes favorably with
ESSC up until $\mu= 0.3$, at which point Infomap places all vertices
in the same community. Interestingly, OSLOM has a peak of performance
around $\mu= 0.6$. This appears to hinge on the fact that the method
measures the strength of a community through assuming that vertices
outside a community are close to the connectivity of the vertex of the
community that has the lowest connectivity for the specified community.
Highly mixed communities tend to favor this similarity, giving OSLOM an
advantage in these cases. Importantly, ESSC performs nearly as well on
networks of disjoint communities \textit{with} background vertices as it
does on these types of networks \textit{without} background (see
the Appendix~\ref{sec:nonbacksims} for nonbackground simulations). On the other
hand, the remaining methods tend to, on average, perform much worse
when background vertices are introduced.

\section{Discussion}

The identification of communities of tightly connected vertices in
networks has proven to be an important tool in the exploratory analysis
and study of a variety of complex connected systems. In this paper we
introduced a means to measure the statistical significance of
connection between a single vertex and any collection of vertices in
undirected networks through a reference distribution derived from the
properties of the conditional configuration model. We introduced and
evaluated a testing based community detection method, ESSC, which
identifies statistically significant communities through the use of
$p$-values derived from this reference distribution. This method
automatically chooses the number of communities and relies only upon
one parameter which guides the false discovery rate of discovered communities.

The ESSC extraction technique directly addresses the importance of
identifying background vertices within a network that need not
necessarily be assigned to identified communities. Given the
heterogeneities of vertex roles in most real-world network data,
identifying background nodes is an important aspect of community
detection. Methods which identify background vertices can help prevent
the noise associated with their connections from polluting the
otherwise significant features among and between communities.

We evaluated ESSC and a number of competing community detection methods
using a variety of quantitative and network-specific validation
measures. We have shown that ESSC is able to capture features of
network data that are relevant to the modeled complex system. For
instance, in the Caltech network study we found that ESSC identified
communities closely matching the dormitory residence of its
individuals; similarly, in the political blog study ESSC identified
communities matching the political affiliation of the bloggers in the
network. Importantly, ESSC identified a moderate amount of background
for each analyzed network in this paper, suggesting potential benefits
to distinguishing background in a network.

Finally, through a series of simulations we have shown that ESSC is
able to successfully capture both overlapping and disjoint community
structure, as well as community structure in networks with background.
In the former scenario, ESSC is competitive with many modern detection
methods, while in the latter we find that ESSC outperforms competing methods.

The development of ESSC relied on undirected, unweighted networks,
however, this can be extended to networks of different structures,
including directed, multilayer and time-varying networks. Understanding
the statistical significance of communities in each of these more
complex network structures requires both theoretical and methodological
work, providing avenues for future research. {This includes comparing
ESSC to the various stochastic block model fitting algorithms and other
permutation-based statistical methods that have been recently developed
over the past few years.} Furthermore, understanding the
consistency properties of the ESSC algorithm is an interesting question
of independent interest which will require recently developed
probabilistic tools.

\begin{appendix}\label{app}
\section{Approximate distribution of \texorpdfstring{$\hat{d}(u:B)$}{$d(u:B)$}}\label{sec:theory}
Here we state and prove Theorem~\ref{prop:binomial} which gives the
approximate law of $\hat{d}_n(u:B)$ on which our algorithm is based in
the large network limit. The result is specific to the conditional
configuration model, which we use as a null network model in order to
find significant community structure.

\begin{pf*}{Proof of Theorem~\ref{prop:binomial}}
Equation \eqref{eq:A} implies that for the number of edges $E_{o,n}$
one has
\[
\int_{\bR} x \,dF_n(x) = \sum
_{k=0}^\infty k \frac{N_k(n)}{n} = 2\frac
{|E_{o,n}|}{n}
\sim\mu,
\]
where $N_k(n)$ is the number of vertices of degree $k$. Thus,
$|E_{o,n}| \sim n \mu/2$.

Now to understand the distribution of $\hat{d}_n(u: B)$, namely, the
number of connections of vertex $u$ to the subset $B$ in $\operatorname{CM}(\vd
_{o,n})$, we use the fact that for constructing the configuration
model, one can start at any vertex and start sequentially attaching the
half-edges of that vertex at random to available\vadjust{\goodbreak} half-edges. We start
with the fixed vertex $u$ and decide the half-edges paired to the
$d_{o,n}(u):=k $ half-edges of vertex $u$. Write $A_1$ for the event
that the first half-edge of vertex $v$ connects to the set $B$ and
write $r_1(B)$ for the probability of this event. Then,
%
%
\begin{equation}
r_1(B) = \frac{\sum_{v\in B} d_{o,n}(v)}{[\sum_{v \in[n]}
d_{o,n}(v)]-1 } = \frac{\sum_{v\in B} d_{o,n}(v)}{2|E_{o,n}|-1}.
\end{equation}
{Now if each half-stub sampled with replacement from the stubs
corresponding to set $B$, then $\hat{d}_n(u:B)$ would exactly
correspond to a Binomial distribution. The main issue to understand is
the effect of sampling without replacement from the half-stubs of $B$,
namely, once a half-stub of $B$ is used by $u$, it cannot be reused. }
In general, for $1\leq i\leq k$, let $A_i$ denote the event that
half-edge $i$ connects to the set $B$ and write $r_i(B)$ for the
conditional probability of $A_i$ conditional on the outcomes of the
first $i-1$ choices. For $i=2$, we claim that uniformly on all outcomes
for the first edge, this conditional probability can be bounded as
%
%
\begin{equation}
\label{eqn-p2} \frac{[\sum_{v\in B} d_{o,n}(v)]-1}{2|E_{o,n}|-2} \leq r_2(B) \leq
\frac{\sum_{v\in B} d_{o,n}(v)}{2|E_{o,n}|-2 }.
\end{equation}
The lower bound arises if the first half-edge of $v$ connected to a
half-edge of $B$, while the upper bound arises if the first half-edge
does not connect to a half-edge emanating from $B$. Arguing analogously
for $1\leq i\leq k$, we find that the conditional probability $r_i(B)$
that the $i$th half-edge of vertex $v$ connects to $B$ is bounded
(uniformly on all choices of the first $i-1$ edges) as
%
%
\begin{equation}
\label{eqn-pi} \frac{[\sum_{v\in B} d_{o,n}(v)]-(i-1)}{2|E_{o,n}|-i } \leq r_i(B) \leq
\frac{\sum_{v\in B} d_{o,n}(v)}{2|E_{o,n}|-i }.
\end{equation}
Recall that $p_n(B) = \sum_{v\in B} d_{o,n}(v)/2|E_{o,n}| $. Since
$|E_{o,n}| \sim n\mu/2$, using \eqref{eqn-pi}, we have
%
%
\begin{equation}
\label{eqn:sup-p} \sup_{1\leq i\leq k}\bigl|r_i(B) -
p_n(B)\bigr| \leq3 \frac{k}{2|E_{o,n}|} + O \biggl( \biggl(
\frac{k}{2|E_{o,n}|} \biggr)^2 \biggr) \to0
\end{equation}
as $n\to\infty$.

Now note that the random variable of interest $\hat{d}_n(u:B)$ can be
expressed as
\[
\hat{d}_n(u:B) = \sum_{i=1}^k
\mathbh{1}\{A_i\}.
\]
Equation \eqref{eqn:sup-p} implies that
\[
d_{\mathrm{TV}}\bigl(\hat{d}_n(u:B), \operatorname{Bin}\bigl(k,
p_n(B)\bigr)\bigr) \to0\qquad \mbox{as } n\to\infty.
\]
This completes the proof.
\end{pf*}

\section{Simulations on disjoint and overlapping community
benchmarks}\label{sec:nonbacksims}
\begin{figure}

\includegraphics{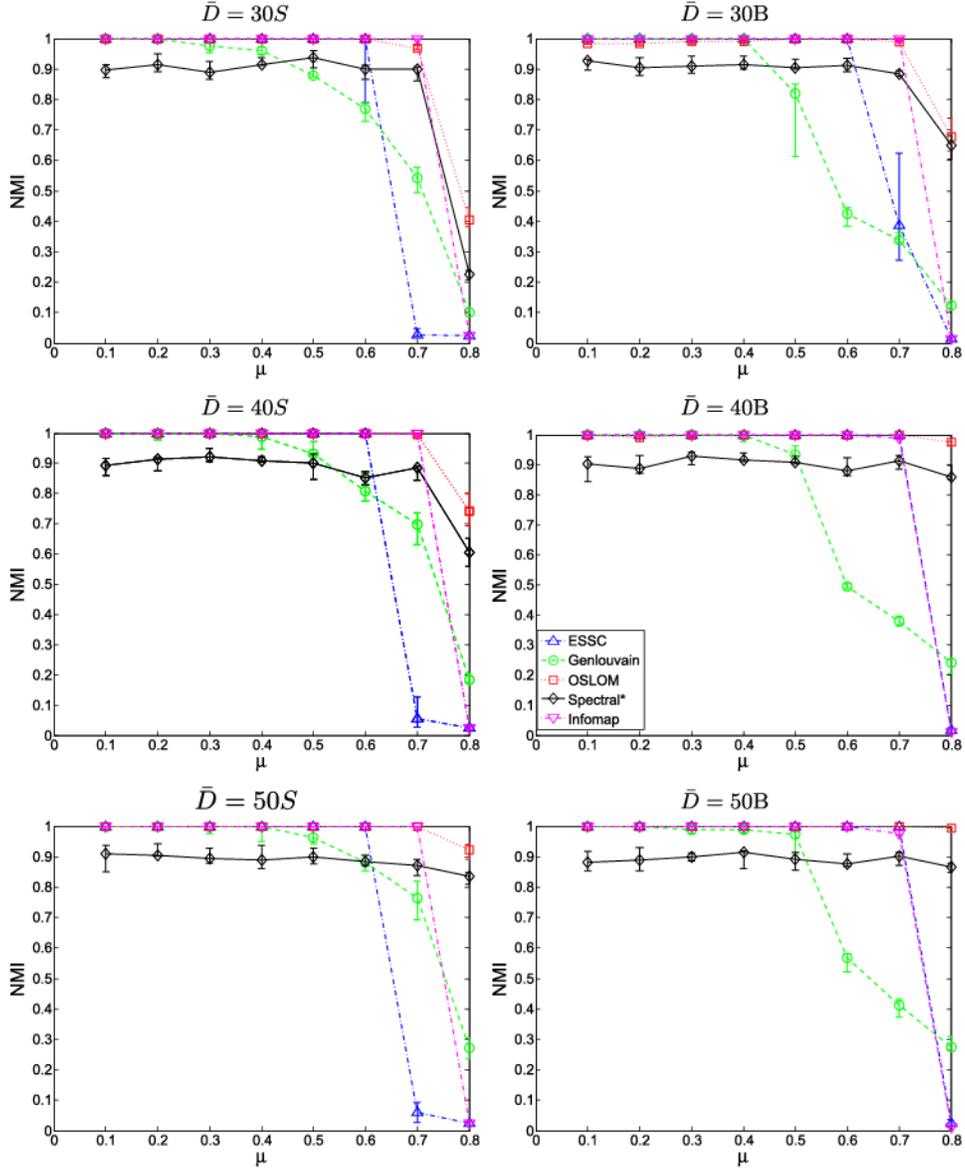}

\caption{The results on the LFR disjoint benchmarks.
Shown are the first, second and third quartile match of
each method over 30 realizations across values of $\mu$.
The degree distribution follows a power law with
exponent $\tau_1 = 2$ with average degree specified
in each plot. *Here, Spectral was given the true number of
communities.}\label{fig:LFR_disjoint} 
\end{figure}

\textit{Disjoint communities}:
LFR benchmarks of size 2000 were simulated with two ranges of community
size, $[10,50]$ (small, S) and $[20,100]$ (big, B), where the community
sizes were derived from a power-law distribution with exponent $\tau_2
= 1$ and with average degree $\bar{D}$ equal to 30, 40 or 50 with
degrees deriving from a power-law distribution with exponent $\tau_1 = 2$.
For each value of $\bar{D}$, networks were generated with values of
$\mu
$ ranging from 0.1 to 0.8 in increments
of 0.1.
Thirty realizations were generated from each set of parameters, and the
resulting networks were input to
the ESSC, GenLouvain, Infomap, OSLOM and Spectral methods.
For Spectral, the parameter $k$ was set to the true number of
communities, thereby providing
Spectral with an advantage over the other methods considered.
Normalized mutual information (NMI) [\citet{lancichinetti2009community}]
was used as a measure
of performance for all methods. The results are summarized in
Figure~\ref{fig:LFR_disjoint}.
%

%

%

ESSC performs well (NMI $\approx1$) for all simulations with mixing
parameter $\mu\leq0.6$. In networks with small communities ($[10,50]$),
ESSC finds no significant communities for extreme values of $\mu$
($\geq0.7$).
In networks with larger communities ($[20,100]$), ESSC identifies
underlying structure when $\mu= 0.7$, and performs
particularly well for dense networks ($\bar{D} = 40,50$).
These results suggests that, when communities are weakly defined, ESSC
performs better when the underlying
communities are large. Overall, ESSC, OSLOM and Infomap performed
ideally when $\mu\leq0.6$.

\begin{figure}

\includegraphics{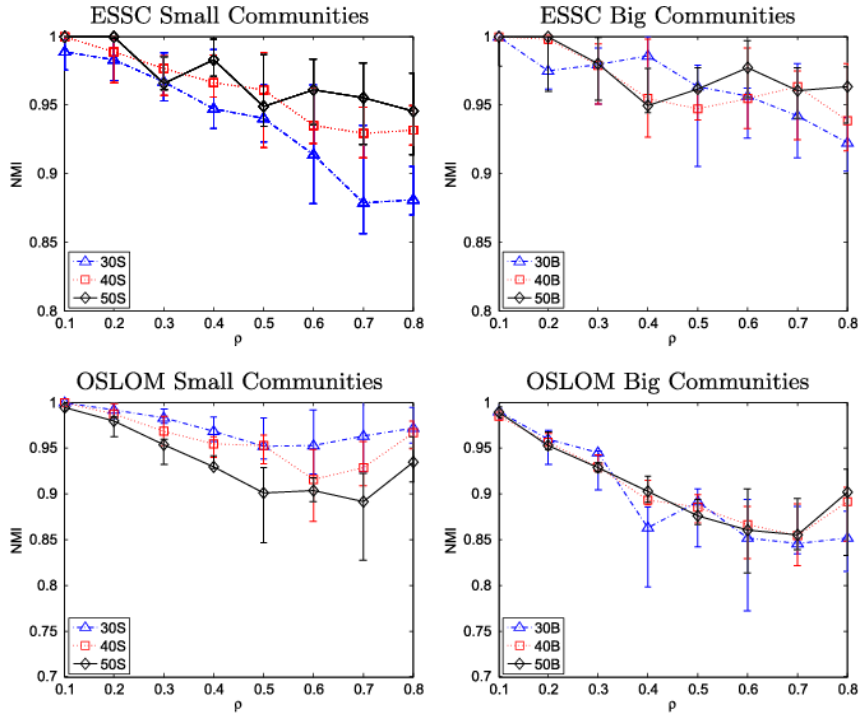}

\caption{The results on the LFR overlapping benchmarks. Shown are the
first, second and third quartile match of each method over 30
realizations across values of $\rho$ at fixed $\mu= 0.3$ for both
small [30--50] and big [50--100] communities. The degree distribution
follows a power law with exponent $\tau_1 = 2$ with average degree
specified by the color of each line.} \label{fig:LFRoverlap}
\end{figure}

\textit{Overlapping communities}:
LFR benchmarks of size 2000 were simulated with two ranges of community
size, $[10,50]$ (small, S)
and $[20,100]$ (big, B), with size distribution following a power law
with exponent $\tau_2 = 1$ and with average degree $\bar{D}$ equal to
30, 40 or 50 where the degree distribution follows a power law with
exponent $\tau_1 = 2$.
For each value of $\bar{D}$, networks were generated with values of
$\rho$ ranging from 0.1 to 0.8
in increments of 0.1. The mixing parameter $\mu$ was set to 0.3.
Thirty realizations were generated from each set of parameters and then
input to ESSC and OSLOM. Once again, the generalized NMI was used to
evaluate the similarity between the detected communities and the true
cover. The results are summarized in Figure~\ref{fig:LFRoverlap}.

From Figure~\ref{fig:LFRoverlap}, we first notice that ESSC performs
competitively with OSLOM in detecting overlapping community structure
across all $\rho$. In networks with small communities (size in
$[10,50]$), the performance of ESSC improves as the density of the
network increases. We also see that ESSC improves when the size of the
communities increases as observed by comparing the left and right
panels of the ESSC results in Figure~\ref{fig:LFRoverlap}. This agrees\vadjust{\eject}
with our observation in the disjoint community study suggesting that
ESSC prefers networks with larger communities.

%
%

%
%

\section{Parameter settings of detection~methods}\label{sec:parameters}
We now describe the exact parameter settings as well as the code used
for all detection methods throughout our real network analysis and
simulation studies in Sections~\ref{sec:realdata}--\ref{sec:simulations}:
\begin{itemize}
\item\textit{ESSC}: We use the MATLAB implementation of the algorithm
provided by the authors at \url
{http://www.unc.edu/\textasciitilde jameswd/research.html}. We set
$\alpha$ to be 0.05
for all real data sets and simulated networks except for the Caltech
Facebook network where we set $\alpha$ to be 0.01.
\item\textit{OSLOM}: We use the C$++$ implementation available at \url
{http://www.oslom.org/software.htm}. For each study we use the default
settings\vadjust{\goodbreak} under an unweighted undirected network with no hierarchy. The
$p$-value threshold is by default set at 0.1. A random seed is used for
its random number generator.
\item\textit{Infomap}: We use the C$++$ implementation available at
\surl
{http://www.\\mapequation.org/code.html}. For each study we use the
default settings of the algorithm for an undirected network. We use a
random positive integer as the seed and run 500 attempts of the
algorithm to partition the network.
\item\textit{$k$-means}: We use the MATLAB implementation of the
algorithm that is available for current MATLAB software. In each study
we choose $k$ according to the network as described throughout the
text. We ran the algorithm over 500 iterations and used a random seed
for initialization.
\item\textit{Spectral clustering}: We use the MATLAB implementation of
the normalized Spectral Clustering algorithm. We choose $k$ according
to the network as described in the text. Again, we ran the algorithm
over 500 iterations and used a random seed for initialization.
\item\textit{GenLouvain}: We use the MATLAB implementation of the
generalized version of Louvain (GenLouvain) from \citet{Jutla2011}. For
the real network analysis, we run the algorithm across a range of
resolution parameters, $\gamma$ ranging from 0.1 to 1.0 (in increments
of 0.1). For each $\gamma$, we look at the number of communities of the
resulting partition and choose $\gamma$ to be the first value for which
the size is stable in terms of being constant across neighboring values
of $\gamma$. In doing so, we chose $\gamma= 0.8$ for the Caltech
Facebook network and $\gamma= 0.3$ for the political blog network. In
the simulation study, we use the randomized version of GenLouvain
(available on the same website) and choose the partition of the highest
modularity across 30 repetitions. In each run, we use the default
resolution parameter $\gamma= 1$. We use a random seed for each run of
the algorithm.

%
%
\begin{table}[t]
\caption{A summary of the communities detected by ESSC across a range
of values of $\alpha$ when run on the Caltech Facebook network. These
statistics are the same as those presented in Section \protect\ref
{sec:realdata}}\label{tab8}
\begin{tabular*}{\textwidth}{@{\extracolsep{\fill}}lcd{3.2}cccd{2.2}c@{}}
\hline
\multicolumn{1}{@{}l}{$\bolds{\alpha}$} & \multicolumn{1}{c}{$\bolds{N_C}$} & \multicolumn{1}{c}{$\bolds{\overline{S}}$} &
\multicolumn{1}{c}{$\bolds{\hat{\sigma}_S}$} & \multicolumn{1}{c}{$\bolds{\overline
{\mathbb{M}}}$} & \multicolumn{1}{c}{$\bolds{\overline{D}_{\mathrm{in}}}$} & \multicolumn{1}{c}{$\bolds{\overline{D}_{\mathrm{out}}}$} & \multicolumn{1}{c@{}}{$\bolds{P_B}$}\\
\hline
0.01 & 7& 78.57 &16.03 &1.03 & 55.76& 15.81&0.30 \\
0.02 & 7& 80.29& 15.52& 1.04& 55.52& 14.97&0.29 \\
0.03 & 7& 82.43& 15.05& 1.05& 55.14& 14.41&0.28 \\
0.04 & 6& 86.67& 12.40& 1.02& 56.34& 17.98&0.33 \\
0.05 & 6& 94.33& 14.02& 1.07& 55.25& 17.33&0.30 \\
0.06 & 6& 95.67& 14.12& 1.07& 54.92& 17.26&0.30 \\
0.07 & 6& 97.33& 14.99& 1.07& 54.58& 16.04&0.28 \\
0.08 & 6& 98.17& 14.93& 1.07& 54.16& 16.93&0.28 \\
0.09 & 8& 110.63& 22.61& 1.28& 52.38& 7.42&0.19 \\
0.10 & 8& 117.13& 31.02& 1.36& 51.95& 9.50&0.19 \\
\hline
\end{tabular*}
\end{table}
%
%
\begin{table}[b]
\tablewidth=220pt
\caption{The Jaccard score of the background vertices distinguished at
each value of $\alpha$ when compared to the background vertices found
with $\alpha= 0.01$. These analyses are done on the Caltech Facebook
presented in Section \protect\ref{sec:realdata}}\label{tab9}
\begin{tabular*}{220pt}{@{\extracolsep{\fill}}lc@{}}
\hline
$\bolds{\alpha}$ & \multicolumn{1}{c@{}}{\textbf{Jaccard score}}\\
\hline
0.01 & 1.00\phantom{00} \\
0.02 & 0.9652 \\
0.03 & 0.9304 \\
0.04 & 0.8015 \\
0.05 & 0.7303\\
0.06 & 0.7116 \\
0.07 & 0.6985 \\
0.08 & 0.7011\\
0.09 & 0.5907 \\
0.10 & 0.6085\\
\hline
\end{tabular*}
\end{table}

\item\textit{ZLZ}: We use the R implementation provided to us by the
author Yunpeng Zhao. We run the tabu search part of the algorithm 1000
iterations for each run. We choose $k$ according to the network as
described in our report. The normalized default score from \citet
{zhao2011community} was used as the objective function to which the
algorithm was run to optimize. A~random seed was set for initialization.
\end{itemize}

%
\section{On the effects of \texorpdfstring{$\alpha$}{alpha}}\label{sec:alpha}
As discussed in the main paper, $\alpha$ is the only tunable parameter
of the ESSC algorithm. The value of $\alpha$ controls the level for
which communities are declared statistically signficant. To get an idea
of how sensitive the algorithm is to this parameter, we run the
algorithm on the first two analyzed data\vadjust{\goodbreak} sets---the Caltech Facebook
network and the political blog network---with values of $\alpha$
between 0.01 and 0.10. We summarize the detected communities using the
statistics of Section~\ref{sec:realdata}.
A summary of results are provided in Tables \ref{tab8} and \ref{tab10}.
The match of the identified communities with those discussed in
the main text are given in Tables \ref{tab9} and \ref{tab11}.
Further, we assess the
similarity of the background vertices from each setting using the
Jaccard score. The match and statistics are shown below. In general,
these statistics suggest\vadjust{\eject} that the communities detected by the ESSC
algorithm are robust in the sense that they are not sensitive to the
choice of $\alpha$.

%
\begin{table}
\caption{A summary of the communities detected by ESSC across a range
of values of $\alpha$ when run on the political blog network. These
statistics are the same as those presented in Section \protect\ref
{sec:realdata}}\label{tab10}
\begin{tabular*}{\textwidth}{@{\extracolsep{\fill}}lcd{3.2}d{3.2}ccd{2.2}c@{}}
\hline
\multicolumn{1}{@{}l}{$\bolds{\alpha}$} & \multicolumn{1}{c}{$\bolds{N_C}$} & \multicolumn{1}{c}{$\bolds{\overline{S}}$} &
\multicolumn{1}{c}{$\bolds{\hat{\sigma}_S}$} & \multicolumn{1}{c}{$\bolds{\overline
{\mathbb{M}}}$} & \multicolumn{1}{c}{$\bolds{\overline{D}_{\mathrm{in}}}$} & \multicolumn{1}{c}{$\bolds{\overline{D}_{\mathrm{out}}}$} & \multicolumn{1}{c@{}}{$\bolds{P_B}$}\\
\hline
0.01 &2&394.5 &54.45 & 1.00& 40.51&3.40 & 0.35\\
0.02 &2&406.5 & 67.18& 1.00& 39.47& 3.27& 0.33\\
0.03 &2&420.0 & 53.74& 1.00& 38.40& 3.07& 0.31\\
0.04 &2 &423.5 & 57.28& 1.00& 38.14& 3.00& 0.31\\
0.05 &2 &448.5 & 75.66& 1.00& 36.30& 2.58& 0.27\\
0.06 &2 &449.5 & 75.66& 1.00& 36.27& 2.45& 0.26\\
0.07 &2 &431.0 & 46.67& 1.00& 37.60& 2.84& 0.29\\
0.08 &3 &528.3 & 146.92& 1.30& 27.37& 24& 0.01\\
0.09 &2 &449.5 & 72.83& 1.00& 36.24& 2.54& 0.26\\
0.10 &3 &323.67 & 249.93&1.02 & 34.39& 2.56& 0.22\\
\hline
\end{tabular*}
\end{table}
%

%
\begin{table}[b]
\tablewidth=200pt
\caption{The Jaccard score of the background vertices distinguished at
each value of $\alpha$ when compared to the background vertices found
with $\alpha= 0.05$. These analyses are done on the political blog
network of Section \protect\ref{sec:realdata}}\label{tab11}
\begin{tabular*}{200pt}{@{\extracolsep{\fill}}lc@{}}
\hline
$\bolds{\alpha}$ & \multicolumn{1}{c@{}}{\textbf{Jaccard score}}\\
\hline
0.01 & 0.7483 \\
0.02 & 0.7922 \\
0.03 & 0.8433 \\
0.04 & 0.8590 \\
0.05 & 1.00\phantom{00} \\
0.06 & 0.9938 \\
0.07 & 0.8843 \\
0.08 & 0.0062 \\
0.09 & 0.9877 \\
0.10 & 0.8277\\
\hline
\end{tabular*}
\end{table}
\end{appendix}


\section*{Acknowledgments}
{We would like to thank the referees, Associate Editor and the Editor
for their constructive suggestions which led to a significant
improvement of the paper.} We would like to thank Mason Porter for
sharing the Caltech Facebook data set that we analyzed in Section~\ref
{sec:caltech}. We would also like to thank Yunpeng Zhao for
contributing his code for the ZLZ extraction algorithm.

\begin{supplement}[id=suppA]
\stitle{Supplemental personal Facebook data set\\}
\slink[doi]{10.1214/14-AOAS760SUPP} 
\sdatatype{.zip}
\sfilename{aoas760\_supp.zip}
\sdescription{We provide the personal Facebook data set as well as
anonymized labels used in the analysis in Section~\ref{sec4.3}
of the manuscript.}
\end{supplement}


%

%

\printaddresses

\begin{thebibliography}{65}

\bibitem[\protect\citeauthoryear{Adamic and
Glance}{2005}]{adamic2005political}
%
\begin{binproceedings}[author]
\bauthor{\bsnm{Adamic},~\bfnm{L.~A.}\binits{L.~A.}} \AND
\bauthor{\bsnm{Glance},~\bfnm{N.}\binits{N.}}
(\byear{2005}).
\btitle{The political blogosphere and the 2004 US election: Divided
they blog}.
In \bbooktitle{Proceedings of the 3rd International Workshop on Link Discovery}
\bpages{36--43}.
\bpublisher{ACM},
\blocation{New York}.
\end{binproceedings}
%
\bptok{imsref}%
\endbibitem

\bibitem[\protect\citeauthoryear{Airoldi, Costa and
Chan}{2013}]{airoldi2013stochastic}
%
\begin{binproceedings}[author]
\bauthor{\bsnm{Airoldi},~\bfnm{Edoardo~M.}\binits{E.~M.}},
\bauthor{\bsnm{Costa},~\bfnm{Thiago~B.}\binits{T.~B.}} \AND
\bauthor{\bsnm{Chan},~\bfnm{Stanley~H.}\binits{S.~H.}}
(\byear{2013}).
\btitle{Stochastic blockmodel approximation of a graphon: Theory and
consistent estimation}.
In \bbooktitle{Advances in Neural Information Processing Systems}
\bpages{692--700}.
\end{binproceedings}
%
\bptok{imsref}%
\endbibitem

\bibitem[\protect\citeauthoryear{Airoldi et~al.}{2008}]{airoldi2008mixed}
%
\begin{barticle}[author]
\bauthor{\bsnm{Airoldi},~\bfnm{Edoardo~M.}\binits{E.~M.}},
\bauthor{\bsnm{Blei},~\bfnm{David~M.}\binits{D.~M.}},
\bauthor{\bsnm{Fienberg},~\bfnm{Stephen~E.}\binits{S.~E.}} \AND
\bauthor{\bsnm{Xing},~\bfnm{Eric~P.}\binits{E.~P.}}
(\byear{2008}).
\btitle{Mixed membership stochastic blockmodels}.
\bjournal{J. Mach. Learn. Res.}
\bvolume{9}
\bpages{1981--2014}.
\end{barticle}
%
\bptok{imsref}%
\endbibitem

\bibitem[\protect\citeauthoryear{Amini et~al.}{2013}]{amini2013pseudo}
%
\begin{barticle}[mr]
\bauthor{\bsnm{Amini},~\bfnm{Arash~A.}\binits{A.~A.}},
\bauthor{\bsnm{Chen},~\bfnm{Aiyou}\binits{A.}},
\bauthor{\bsnm{Bickel},~\bfnm{Peter~J.}\binits{P.~J.}} \AND
\bauthor{\bsnm{Levina},~\bfnm{Elizaveta}\binits{E.}}
(\byear{2013}).
\btitle{Pseudo-likelihood methods for community detection in large
sparse networks}.
\bjournal{Ann. Statist.}
\bvolume{41}
\bpages{2097--2122}.
\bid{doi={10.1214/13-AOS1138}, issn={0090-5364}, mr={3127859}}
\end{barticle}
%
\bptok{imsref}%
\endbibitem

\bibitem[\protect\citeauthoryear{Ball, Karrer and
Newman}{2011}]{ball2011efficient}
%
\begin{barticle}[author]
\bauthor{\bsnm{Ball},~\bfnm{B.}\binits{B.}},
\bauthor{\bsnm{Karrer},~\bfnm{B.}\binits{B.}} \AND
\bauthor{\bsnm{Newman},~\bfnm{M.~E.~J.}\binits{M.~E.~J.}}
(\byear{2011}).
\btitle{Efficient and principled method for detecting communities in networks}.
\bjournal{Phys. Rev. E (3)}
\bvolume{84}
\bpages{036103}.
\end{barticle}
%
\bptok{imsref}%
\endbibitem

\bibitem[\protect\citeauthoryear{Barab{\'a}si and
Albert}{1999}]{barabasi1999emergence}
%
\begin{barticle}[mr]
\bauthor{\bsnm{Barab{\'a}si},~\bfnm{Albert-L{\'a}szl{\'o}}\binits
{A.-L.}} \AND
\bauthor{\bsnm{Albert},~\bfnm{R{\'e}ka}\binits{R.}}
(\byear{1999}).
\btitle{Emergence of scaling in random networks}.
\bjournal{Science}
\bvolume{286}
\bpages{509--512}.
\bid{doi={10.1126/science.286.5439.509}, issn={0036-8075}, mr={2091634}}
\end{barticle}
%
\bptok{imsref}%
\endbibitem

\bibitem[\protect\citeauthoryear{Bassett et~al.}{2011}]{Bassett2011}
%
\begin{barticle}[author]
\bauthor{\bsnm{Bassett},~\bfnm{Danielle~S.}\binits{D.~S.}},
\bauthor{\bsnm{Wymbs},~\bfnm{Nicholas~F.}\binits{N.~F.}},
\bauthor{\bsnm{Porter},~\bfnm{Mason~A.}\binits{M.~A.}},
\bauthor{\bsnm{Mucha},~\bfnm{Peter~J.}\binits{P.~J.}},
\bauthor{\bsnm{Carlson},~\bfnm{Jean~M.}\binits{J.~M.}} \AND
\bauthor{\bsnm{Grafton},~\bfnm{Scott~T.}\binits{S.~T.}}
(\byear{2011}).
\btitle{Dynamic reconfiguration of human brain networks during learning}.
\bjournal{Proc. Natl. Acad. Sci. USA}
\bvolume{108}
\bpages{7641--7646}.
\end{barticle}
%
\bptok{imsref}%
\endbibitem

\bibitem[\protect\citeauthoryear{Bender and
Canfield}{1978}]{bender1978asymptotic}
%
\begin{barticle}[mr]
\bauthor{\bsnm{Bender},~\bfnm{Edward~A.}\binits{E.~A.}} \AND
\bauthor{\bsnm{Canfield},~\bfnm{E.~Rodney}\binits{E.~R.}}
(\byear{1978}).
\btitle{The asymptotic number of labeled graphs with given degree sequences}.
\bjournal{J. Combin. Theory Ser. A}
\bvolume{24}
\bpages{296--307}.
\bid{mr={0505796}}
\end{barticle}
%
\bptok{imsref}%
\endbibitem

\bibitem[\protect\citeauthoryear{Benjamini and
Hochberg}{1995}]{benjamini1995controlling}
%
\begin{barticle}[mr]
\bauthor{\bsnm{Benjamini},~\bfnm{Yoav}\binits{Y.}} \AND
\bauthor{\bsnm{Hochberg},~\bfnm{Yosef}\binits{Y.}}
(\byear{1995}).
\btitle{Controlling the false discovery rate: A practical and powerful
approach to multiple testing}.
\bjournal{J. R. Stat. Soc. Ser. B Stat. Methodol.}
\bvolume{57}
\bpages{289--300}.
\bid{issn={0035-9246}, mr={1325392}}
\end{barticle}
%
\bptok{imsref}%
\endbibitem

\bibitem[\protect\citeauthoryear{Bickel and
Chen}{2009}]{bickel2009nonparametric}
%
\begin{barticle}[author]
\bauthor{\bsnm{Bickel},~\bfnm{P.~J.}\binits{P.~J.}} \AND
\bauthor{\bsnm{Chen},~\bfnm{A.}\binits{A.}}
(\byear{2009}).
\btitle{A nonparametric view of network models and Newman--Girvan and
other modularities}.
\bjournal{Proc. Natl. Acad. Sci. USA}
\bvolume{106}
\bpages{21068--21073}.
\end{barticle}
%
\bptok{imsref}%
\endbibitem

\bibitem[\protect\citeauthoryear{Blondel et~al.}{2008}]{blondel2008fast}
%
\begin{barticle}[author]
\bauthor{\bsnm{Blondel},~\bfnm{Vi.~D.}\binits{Vi.~D.}},
\bauthor{\bsnm{Guillaume},~\bfnm{J.~L.}\binits{J.~L.}},
\bauthor{\bsnm{Lambiotte},~\bfnm{R.}\binits{R.}} \AND
\bauthor{\bsnm{Lefebvre},~\bfnm{E.}\binits{E.}}
(\byear{2008}).
\btitle{Fast unfolding of communities in large networks}.
\bjournal{J. Stat. Mech. Theory Exp.}
\bvolume{2008}
\bpages{P10008}.
\end{barticle}
%
\bptok{imsref}%
\endbibitem

\bibitem[\protect\citeauthoryear{Bollob{\'a}s}{1979}]{bollobas1979probabilistic}
%
\begin{bmisc}[author]
\bauthor{\bsnm{Bollob{\'a}s},~\bfnm{B.}\binits{B.}} 
(\byear{1979}).
\bhowpublished{A probabilistic proof of an asymptotic formula for the number
of labelled regular graphs.
Aarhus Universitet}.
\end{bmisc}
%
\bptok{imsref}%
\endbibitem

\bibitem[\protect\citeauthoryear{Clauset, Moore and
Newman}{2008}]{clauset2008hierarchical}
%
\begin{barticle}[author]
\bauthor{\bsnm{Clauset},~\bfnm{Aaron}\binits{A.}},
\bauthor{\bsnm{Moore},~\bfnm{Cristopher}\binits{C.}} \AND
\bauthor{\bsnm{Newman},~\bfnm{Mark~EJ}\binits{M.~E.}}
(\byear{2008}).
\btitle{Hierarchical structure and the prediction of missing links in
networks}.
\bjournal{Nature}
\bvolume{453}
\bpages{98--101}.
\end{barticle}
%
\bptok{imsref}%
\endbibitem

\bibitem[\protect\citeauthoryear{Clauset, Newman and
Moore}{2004}]{clauset2004finding}
%
\begin{barticle}[author]
\bauthor{\bsnm{Clauset},~\bfnm{A.}\binits{A.}},
\bauthor{\bsnm{Newman},~\bfnm{M.~E.~J.}\binits{M.~E.~J.}} \AND
\bauthor{\bsnm{Moore},~\bfnm{C.}\binits{C.}}
(\byear{2004}).
\btitle{Finding community structure in very large networks}.
\bjournal{Phys. Rev. E (3)}
\bvolume{70}
\bpages{066111}.
\end{barticle}
%
\bptok{imsref}%
\endbibitem

\bibitem[\protect\citeauthoryear{Decelle et~al.}{2011}]{decelle2011inference}
%
\begin{barticle}[pbm]
\bauthor{\bsnm{Decelle},~\bfnm{Aurelien}\binits{A.}},
\bauthor{\bsnm{Krzakala},~\bfnm{Florent}\binits{F.}},
\bauthor{\bsnm{Moore},~\bfnm{Cristopher}\binits{C.}} \AND
\bauthor{\bsnm{Zdeborov{\'{a}}},~\bfnm{Lenka}\binits{L.}}
(\byear{2011}).
\btitle{Inference and phase transitions in the detection of modules in
sparse networks}.
\bjournal{Phys. Rev. Lett.}
\bvolume{107}
\bpages{065701}.
\bid{issn={1079-7114}, pmid={21902340}}
\end{barticle}
%
\bptok{imsref}%
\endbibitem

\bibitem[\protect\citeauthoryear{Erd{\H{o}}s and R{\'
e}nyi}{1960}]{erdHos1960evolution}
%
\begin{barticle}[mr]
\bauthor{\bsnm{Erd{\H{o}}s},~\bfnm{P.}\binits{P.}} \AND
\bauthor{\bsnm{R{\'e}nyi},~\bfnm{A.}\binits{A.}}
(\byear{1960}).
\btitle{On the evolution of random graphs}.
\bjournal{Magyar Tud. Akad. Mat. Kutat\'o Int. K\"ozl.}
\bvolume{5}
\bpages{17--61}.
\bid{mr={0125031}}
\end{barticle}
%
\bptok{imsref}%
\endbibitem

\bibitem[\protect\citeauthoryear{Ester et~al.}{1996}]{ester1996density}
%
\begin{binproceedings}[author]
\bauthor{\bsnm{Ester},~\bfnm{Martin}\binits{M.}},
\bauthor{\bsnm{Kriegel},~\bfnm{Hans-Peter}\binits{H.-P.}},
\bauthor{\bsnm{Sander},~\bfnm{J{\"o}rg}\binits{J.}} \AND
\bauthor{\bsnm{Xu},~\bfnm{Xiaowei}\binits{X.}}
(\byear{1996}).
\btitle{A density-based algorithm for discovering clusters in large
spatial databases with noise.}
In \bbooktitle{KDD}
\bvolume{96}
\bpages{226--231}.
\end{binproceedings}
%
\bptok{imsref}%
\endbibitem

\bibitem[\protect\citeauthoryear{Fortunato}{2010}]{fortunato2010community}
%
\begin{barticle}[mr]
\bauthor{\bsnm{Fortunato},~\bfnm{Santo}\binits{S.}}
(\byear{2010}).
\btitle{Community detection in graphs}.
\bjournal{Phys. Rep.}
\bvolume{486}
\bpages{75--174}.
\bid{doi={10.1016/j.physrep.2009.11.002}, issn={0370-1573}, mr={2580414}}
\end{barticle}
%
\bptok{imsref}%
\endbibitem

\bibitem[\protect\citeauthoryear{Freund and Schapire}{1997}]{adaboost1997}
%
\begin{barticle}[mr]
\bauthor{\bsnm{Freund},~\bfnm{Yoav}\binits{Y.}} \AND
\bauthor{\bsnm{Schapire},~\bfnm{Robert~E.}\binits{R.~E.}}
(\byear{1997}).
\btitle{A decision-theoretic generalization of on-line learning and an
application to boosting}.
\bjournal{J. Comput. System Sci.}
\bvolume{55}
\bpages{119--139}.
\bid{doi={10.1006/jcss.1997.1504}, issn={0022-0000}, mr={1473055}}
\bptnote{check year}%
\end{barticle}
%
\bptok{imsref}%
\endbibitem

\bibitem[\protect\citeauthoryear{Girvan and
Newman}{2002}]{girvan2002community}
%
\begin{barticle}[mr]
\bauthor{\bsnm{Girvan},~\bfnm{M.}\binits{M.}} \AND
\bauthor{\bsnm{Newman},~\bfnm{M.~E.~J.}\binits{M.~E.~J.}}
(\byear{2002}).
\btitle{Community structure in social and biological networks}.
\bjournal{Proc. Natl. Acad. Sci. USA}
\bvolume{99}
\bpages{7821--7826 (electronic)}.
\bid{doi={10.1073/pnas.122653799}, issn={1091-6490}, mr={1908073}}
\end{barticle}
%
\bptok{imsref}%
\endbibitem

\bibitem[\protect\citeauthoryear{Glover}{1989}]{glover1989tabu}
%
\begin{barticle}[author]
\bauthor{\bsnm{Glover},~\bfnm{F.}\binits{F.}}
(\byear{1989}).
\btitle{Tabu search---part I}.
\bjournal{ORSA Journal on Computing}
\bvolume{1}
\bpages{190--206}.
\end{barticle}
%
\bptok{imsref}%
\endbibitem

\bibitem[\protect\citeauthoryear{Goldberg and Tarjan}{1988}]{goldberg1988new}
%
\begin{barticle}[mr]
\bauthor{\bsnm{Goldberg},~\bfnm{Andrew~V.}\binits{A.~V.}} \AND
\bauthor{\bsnm{Tarjan},~\bfnm{Robert~E.}\binits{R.~E.}}
(\byear{1988}).
\btitle{A new approach to the maximum-flow problem}.
\bjournal{J.~Assoc. Comput. Mach.}
\bvolume{35}
\bpages{921--940}.
\bid{doi={10.1145/48014.61051}, issn={0004-5411}, mr={1072405}}
\end{barticle}
%
\bptok{imsref}%
\endbibitem

\bibitem[\protect\citeauthoryear{Goldenberg
et~al.}{2010}]{goldenberg2010survey}
%
\begin{barticle}[author]
\bauthor{\bsnm{Goldenberg},~\bfnm{Anna}\binits{A.}},
\bauthor{\bsnm{Zheng},~\bfnm{Alice~X.}\binits{A.~X.}},
\bauthor{\bsnm{Fienberg},~\bfnm{Stephen~E.}\binits{S.~E.}} \AND
\bauthor{\bsnm{Airoldi},~\bfnm{Edoardo~M.}\binits{E.~M.}}
(\byear{2010}).
\btitle{A survey of statistical network models}.
\bjournal{Foundations and Trends in Machine Learning}
\bvolume{2}
\bpages{129--233}.
\end{barticle}
%
\bptok{imsref}%
\endbibitem

\bibitem[\protect\citeauthoryear{Greene, Doyle and
Cunningham}{2010}]{Greene2010}
%
\begin{bincollection}[author]
\bauthor{\bsnm{Greene},~\bfnm{D.}\binits{D.}},
\bauthor{\bsnm{Doyle},~\bfnm{D.}\binits{D.}} \AND
\bauthor{\bsnm{Cunningham},~\bfnm{P.}\binits{P.}}
(\byear{2010}).
\btitle{Tracking the evolution of communities in dynamic social networks}.
In \bbooktitle{International Conference on Advances in Social Networks
Analysis and Mining (ASONAM)}
\bpages{176--183}.
\bpublisher{Springer},
\blocation{New York}.
\end{bincollection}
%
\bptok{imsref}%
\endbibitem

\bibitem[\protect\citeauthoryear{Handcock, Raftery and
Tantrum}{2007}]{handcock2007model}
%
\begin{barticle}[mr]
\bauthor{\bsnm{Handcock},~\bfnm{Mark~S.}\binits{M.~S.}},
\bauthor{\bsnm{Raftery},~\bfnm{Adrian~E.}\binits{A.~E.}} \AND
\bauthor{\bsnm{Tantrum},~\bfnm{Jeremy~M.}\binits{J.~M.}}
(\byear{2007}).
\btitle{Model-based clustering for social networks}.
\bjournal{J. Roy. Statist. Soc. Ser. A}
\bvolume{170}
\bpages{301--354}.
\bid{doi={10.1111/j.1467-985X.2007.00471.x}, issn={0964-1998}, mr={2364300}}
\end{barticle}
%
\bptok{imsref}%
\endbibitem

\bibitem[\protect\citeauthoryear{Hastie, Tibshirani and
Friedman}{2001}]{friedman2001elements}
%
\begin{bbook}[mr]
\bauthor{\bsnm{Hastie},~\bfnm{Trevor}\binits{T.}},
\bauthor{\bsnm{Tibshirani},~\bfnm{Robert}\binits{R.}} \AND
\bauthor{\bsnm{Friedman},~\bfnm{Jerome}\binits{J.}}
(\byear{2001}).
\btitle{The Elements of Statistical Learning: Data Mining, Inference,
and Prediction}.
\bpublisher{Springer},
\blocation{New York}.
\bid{doi={10.1007/978-0-387-21606-5}, mr={1851606}}
\end{bbook}
%
\bptok{imsref}%
\endbibitem

\bibitem[\protect\citeauthoryear{Hinneburg and
Keim}{1998}]{hinneburg1998efficient}
%
\begin{binproceedings}[author]
\bauthor{\bsnm{Hinneburg},~\bfnm{Alexander}\binits{A.}} \AND
\bauthor{\bsnm{Keim},~\bfnm{Daniel~A.}\binits{D.~A.}}
(\byear{1998}).
\btitle{An efficient approach to clustering in large multimedia
databases with noise}.
In \bbooktitle{KDD, 1998}
\bpages{58--65}.
\end{binproceedings}
%
\bptok{imsref}%
\endbibitem

\bibitem[\protect\citeauthoryear{Hoff, Raftery and
Handcock}{2002}]{hoff2002latent}
%
\begin{barticle}[mr]
\bauthor{\bsnm{Hoff},~\bfnm{Peter~D.}\binits{P.~D.}},
\bauthor{\bsnm{Raftery},~\bfnm{Adrian~E.}\binits{A.~E.}} \AND
\bauthor{\bsnm{Handcock},~\bfnm{Mark~S.}\binits{M.~S.}}
(\byear{2002}).
\btitle{Latent space approaches to social network analysis}.
\bjournal{J. Amer. Statist. Assoc.}
\bvolume{97}
\bpages{1090--1098}.
\bid{doi={10.1198/016214502388618906}, issn={0162-1459}, mr={1951262}}
\end{barticle}
%
\bptok{imsref}%
\endbibitem

\bibitem[\protect\citeauthoryear{Holland, Laskey and
Leinhardt}{1983}]{holland1983stochastic}
%
\begin{barticle}[mr]
\bauthor{\bsnm{Holland},~\bfnm{Paul~W.}\binits{P.~W.}},
\bauthor{\bsnm{Laskey},~\bfnm{Kathryn~Blackmond}\binits{K.~B.}}
\AND
\bauthor{\bsnm{Leinhardt},~\bfnm{Samuel}\binits{S.}}
(\byear{1983}).
\btitle{Stochastic blockmodels: First steps}.
\bjournal{Social Networks}
\bvolume{5}
\bpages{109--137}.
\bid{doi={10.1016/0378-8733(83)90021-7}, issn={0378-8733}, mr={0718088}}
\end{barticle}
%
\bptok{imsref}%
\endbibitem

\bibitem[\protect\citeauthoryear{Jutla, Jeub and
Mucha}{2011/2012}]{Jutla2011}
%
\begin{bmisc}[author]
\bauthor{\bsnm{Jutla},~\bfnm{I.~S.}\binits{I.~S.}},
\bauthor{\bsnm{Jeub},~\bfnm{L.~G.~S.}\binits{L.~G.~S.}} \AND
\bauthor{\bsnm{Mucha},~\bfnm{P.~J.}\binits{P.~J.}}
(\byear{2011/2012}).
\bhowpublished{A generalized {L}ouvain method for community detection
implemented in \textsc{Matlab}.
Available at \url{http://netwiki.amath.unc.edu/GenLouvain}}.
\end{bmisc}
%
\bptok{imsref}%
\endbibitem

\bibitem[\protect\citeauthoryear{Krzakala
et~al.}{2013}]{krzakala2013spectral}
%
\begin{bmisc}[author]
\bauthor{\bsnm{Krzakala},~\bfnm{Florent}\binits{F.}},
\bauthor{\bsnm{Moore},~\bfnm{Cristopher}\binits{C.}},
\bauthor{\bsnm{Mossel},~\bfnm{Elchanan}\binits{E.}},
\bauthor{\bsnm{Neeman},~\bfnm{Joe}\binits{J.}},
\bauthor{\bsnm{Sly},~\bfnm{Allan}\binits{A.}},
\bauthor{\bsnm{Zdeborov{\'a}},~\bfnm{Lenka}\binits{L.}} \AND
\bauthor{\bsnm{Zhang},~\bfnm{Pan}\binits{P.}}
(\byear{2013}).
\bhowpublished{Spectral redemption: Clustering sparse networks.
Preprint. Available at \arxivurl{arXiv:1306.5550}.}
\end{bmisc}
%
\bptok{imsref}%
\endbibitem

\bibitem[\protect\citeauthoryear{Lancichinetti and Fortunato}{2009a}]{lancichinetti2009benchmarks}
%
\begin{barticle}[author]
\bauthor{\bsnm{Lancichinetti},~\bfnm{A.}\binits{A.}} \AND
\bauthor{\bsnm{Fortunato},~\bfnm{S.}\binits{S.}}
(\byear{2009}a).
\btitle{Benchmarks for testing community detection algorithms on
directed and weighted graphs with overlapping communities}.
\bjournal{Phys. Rev. E (3)}
\bvolume{80}
\bpages{016118}.
\end{barticle}
%
\bptok{imsref}%
\endbibitem

\bibitem[\protect\citeauthoryear{Lancichinetti and
Fortunato}{2009b}]{lancichinetti2009community}
%
\begin{barticle}[author]
\bauthor{\bsnm{Lancichinetti},~\bfnm{A.}\binits{A.}} \AND
\bauthor{\bsnm{Fortunato},~\bfnm{S.}\binits{S.}}
(\byear{2009}b).
\btitle{Community detection algorithms: A comparative analysis}.
\bjournal{Phys. Rev. E (3)}
\bvolume{80}
\bpages{056117}.
\end{barticle}
%
\bptok{imsref}%
\endbibitem

\bibitem[\protect\citeauthoryear{Lancichinetti, Fortunato and Kert{\'
e}sz}{2009}]{lancichinetti2009detecting}
%
\begin{barticle}[author]
\bauthor{\bsnm{Lancichinetti},~\bfnm{Andrea}\binits{A.}},
\bauthor{\bsnm{Fortunato},~\bfnm{Santo}\binits{S.}} \AND
\bauthor{\bsnm{Kert{\'e}sz},~\bfnm{J{\'a}nos}\binits{J.}}
(\byear{2009}).
\btitle{Detecting the overlapping and hierarchical community structure
in complex networks}.
\bjournal{New J. Phys.}
\bvolume{11}
\bpages{033015}.
\end{barticle}
%
\bptok{imsref}%
\endbibitem

\bibitem[\protect\citeauthoryear{Lancichinetti
et~al.}{2011}]{lancichinetti2011finding}
%
\begin{barticle}[author]
\bauthor{\bsnm{Lancichinetti},~\bfnm{A.}\binits{A.}},
\bauthor{\bsnm{Radicchi},~\bfnm{F.}\binits{F.}},
\bauthor{\bsnm{Ramasco},~\bfnm{J.~J.}\binits{J.~J.}} \AND
\bauthor{\bsnm{Fortunato},~\bfnm{S.}\binits{S.}}
(\byear{2011}).
\btitle{Finding statistically significant communities in networks}.
\bjournal{PloS One}
\bvolume{6}
\bpages{e18961}.
\end{barticle}
%
\bptok{imsref}%
\endbibitem

\bibitem[\protect\citeauthoryear{Lee and
Cunningham}{2013}]{lee2013benchmarking}
%
\begin{bmisc}[author]
\bauthor{\bsnm{Lee},~\bfnm{Conrad}\binits{C.}} \AND
\bauthor{\bsnm{Cunningham},~\bfnm{P{\'a}draig}\binits{P.}}
(\byear{2013}).
\bhowpublished{Benchmarking community detection methods on social
media data.
Preprint. Available at \arxivurl{arXiv:1302.0739}.}
\end{bmisc}
%
\bptok{imsref}%
\endbibitem

\bibitem[\protect\citeauthoryear{Leskovec et~al.}{2009}]{leskovecENRON}
%
\begin{barticle}[mr]
\bauthor{\bsnm{Leskovec},~\bfnm{Jure}\binits{J.}},
\bauthor{\bsnm{Lang},~\bfnm{Kevin~J.}\binits{K.~J.}},
\bauthor{\bsnm{Dasgupta},~\bfnm{Anirban}\binits{A.}} \AND
\bauthor{\bsnm{Mahoney},~\bfnm{Michael~W.}\binits{M.~W.}}
(\byear{2009}).
\btitle{Community structure in large networks: {N}atural cluster sizes
and the absence of large well-defined clusters}.
\bjournal{Internet Math.}
\bvolume{6}
\bpages{29--123}.
\bid{issn={1542-7951}, mr={2736090}}
\end{barticle}
%
\bptok{imsref}%
\endbibitem

\bibitem[\protect\citeauthoryear{Lewis et~al.}{2010}]{Lewis2010}
%
\begin{barticle}[author]
\bauthor{\bsnm{Lewis},~\bfnm{Anna~CF}\binits{A.~C.}},
\bauthor{\bsnm{Jones},~\bfnm{Nick~S.}\binits{N.~S.}},
\bauthor{\bsnm{Porter},~\bfnm{Mason~A.}\binits{M.~A.}} \AND
\bauthor{\bsnm{Deane},~\bfnm{Charlotte~M.}\binits{C.~M.}}
(\byear{2010}).
\btitle{The function of communities in protein interaction networks at
multiple scales}.
\bjournal{BMC Systems Biology}
\bvolume{4}
\bpages{1--14}.
\end{barticle}
%
\bptok{imsref}%
\endbibitem

\bibitem[\protect\citeauthoryear{M{\'e}zard and
Montanari}{2009}]{mezard2009information}
%
\begin{bbook}[mr]
\bauthor{\bsnm{M{\'e}zard},~\bfnm{Marc}\binits{M.}} \AND
\bauthor{\bsnm{Montanari},~\bfnm{Andrea}\binits{A.}}
(\byear{2009}).
\btitle{Information, Physics, and Computation}.
\bpublisher{Oxford Univ. Press},
\blocation{Oxford}.
\bid{doi={10.1093/acprof:oso/9780198570837.001.0001}, mr={2518205}}
\end{bbook}
%
\bptok{imsref}%
\endbibitem

\bibitem[\protect\citeauthoryear{Miritello, Moro and
Lara}{2011}]{Miritello2011}
%
\begin{barticle}[author]
\bauthor{\bsnm{Miritello},~\bfnm{Giovanna}\binits{G.}},
\bauthor{\bsnm{Moro},~\bfnm{Esteban}\binits{E.}} \AND
\bauthor{\bsnm{Lara},~\bfnm{Rub{\'{e}}n}\binits{R.}}
(\byear{2011}).
\btitle{Dynamical strength of social ties in information spreading}.
\bjournal{Phys. Rev. E (3)}
\bvolume{83}
\bpages{045102}.
\end{barticle}
%
\bptok{imsref}%
\endbibitem

\bibitem[\protect\citeauthoryear{Molloy and Reed}{1995}]{molloy1995critical}
%
\begin{barticle}[mr]
\bauthor{\bsnm{Molloy},~\bfnm{Michael}\binits{M.}} \AND
\bauthor{\bsnm{Reed},~\bfnm{Bruce}\binits{B.}}
(\byear{1995}).
\btitle{A critical point for random graphs with a given degree sequence}.
\bjournal{Random Structures Algorithms}
\bvolume{6}
\bpages{161--179}.
\end{barticle}
%
\bptok{imsref}%
\endbibitem

\bibitem[\protect\citeauthoryear{Mucha et~al.}{2010}]{mucha2010community}
%
\begin{barticle}[mr]
\bauthor{\bsnm{Mucha},~\bfnm{Peter~J.}\binits{P.~J.}},
\bauthor{\bsnm{Richardson},~\bfnm{Thomas}\binits{T.}},
\bauthor{\bsnm{Macon},~\bfnm{Kevin}\binits{K.}},
\bauthor{\bsnm{Porter},~\bfnm{Mason~A.}\binits{M.~A.}} \AND
\bauthor{\bsnm{Onnela},~\bfnm{Jukka-Pekka}\binits{J.-P.}}
(\byear{2010}).
\btitle{Community structure in time-dependent, multiscale, and
multiplex networks}.
\bjournal{Science}
\bvolume{328}
\bpages{876--878}.
\bid{doi={10.1126/science.1184819}, issn={0036-8075}, mr={2662590}}
\end{barticle}
%
\bptok{imsref}%
\endbibitem

\bibitem[\protect\citeauthoryear{Muhammad and
Van~Laerhoven}{2013}]{Muhammad2013}
%
\begin{bincollection}[author]
\bauthor{\bsnm{Muhammad},~\bfnm{S.~A.}\binits{S.~A.}} \AND
\bauthor{\bsnm{Van~Laerhoven},~\bfnm{K.}\binits{K.}}
(\byear{2013}).
\btitle{Quantitative analysis of community detection methods for
longitudinal mobile data}.
In \bbooktitle{International Conference on Social Intelligence and
Technology (SOCIETY)}
\bpages{47--56}.
\bpublisher{Springer},
\blocation{New York}.
\end{bincollection}
%
\bptok{imsref}%
\endbibitem

\bibitem[\protect\citeauthoryear{Newman}{2006}]{newman2006modularity}
%
\begin{barticle}[author]
\bauthor{\bsnm{Newman},~\bfnm{M.~E.~J.}\binits{M.~E.~J.}}
(\byear{2006}).
\btitle{Modularity and community structure in networks}.
\bjournal{Proc. Natl. Acad. Sci. USA}
\bvolume{103}
\bpages{8577--8582}.
\end{barticle}
%
\bptok{imsref}%
\endbibitem

\bibitem[\protect\citeauthoryear{Newman and Girvan}{2004}]{newman2004finding}
%
\begin{barticle}[author]
\bauthor{\bsnm{Newman},~\bfnm{M.~E.~J.}\binits{M.~E.~J.}} \AND
\bauthor{\bsnm{Girvan},~\bfnm{M.}\binits{M.}}
(\byear{2004}).
\btitle{Finding and evaluating community structure in networks}.
\bjournal{Phys. Rev. E (3)}
\bvolume{69}
\bpages{026113}.
\end{barticle}
%
\bptok{imsref}%
\endbibitem

\bibitem[\protect\citeauthoryear{Ng, Jordan and Weiss}{2002}]{ng2002spectral}
%
\begin{barticle}[author]
\bauthor{\bsnm{Ng},~\bfnm{A.~Y.}\binits{A.~Y.}},
\bauthor{\bsnm{Jordan},~\bfnm{M.~I.}\binits{M.~I.}} \AND
\bauthor{\bsnm{Weiss},~\bfnm{Y.}\binits{Y.}}
(\byear{2002}).
\btitle{On spectral clustering: Analysis and an algorithm}.
\bjournal{Adv. Neural Inf. Process. Syst.}
\bvolume{2}
\bpages{849--856}.
\end{barticle}
%
\bptok{imsref}%
\endbibitem

\bibitem[\protect\citeauthoryear{Nowicki and
Snijders}{2001}]{nowicki2001estimation}
%
\begin{barticle}[mr]
\bauthor{\bsnm{Nowicki},~\bfnm{Krzysztof}\binits{K.}} \AND
\bauthor{\bsnm{Snijders},~\bfnm{Tom~A.~B.}\binits{T.~A.~B.}}
(\byear{2001}).
\btitle{Estimation and prediction for stochastic blockstructures}.
\bjournal{J. Amer. Statist. Assoc.}
\bvolume{96}
\bpages{1077--1087}.
\bid{doi={10.1198/016214501753208735}, issn={0162-1459}, mr={1947255}}
\end{barticle}
%
\bptok{imsref}%
\endbibitem

\bibitem[\protect\citeauthoryear{Olhede and Wolfe}{2013}]{olhede2013network}
%
\begin{bmisc}[author]
\bauthor{\bsnm{Olhede},~\bfnm{Sofia~C.}\binits{S.~C.}} \AND
\bauthor{\bsnm{Wolfe},~\bfnm{Patrick~J.}\binits{P.~J.}}
(\byear{2013}).
\bhowpublished{Network histograms and universality of blockmodel approximation.
Preprint. Available at \arxivurl{arXiv:1312.5306}.}
\end{bmisc}
%
\bptok{imsref}%
\endbibitem

\bibitem[\protect\citeauthoryear{Onnela et~al.}{2011}]{Onnela2011}
%
\begin{barticle}[pbm]
\bauthor{\bsnm{Onnela},~\bfnm{Jukka-Pekka}\binits{J.-P.}},
\bauthor{\bsnm{Arbesman},~\bfnm{Samuel}\binits{S.}},
\bauthor{\bsnm{Gonz{\'{a}}lez},~\bfnm{Marta~C.}\binits{M.~C.}},
\bauthor{\bsnm{Barab{\'{a}}si},~\bfnm{Albert-L{\'{a}}szl{\'
{o}}}\binits{A.-L.}} \AND
\bauthor{\bsnm{Christakis},~\bfnm{Nicholas~A.}\binits{N.~A.}}
(\byear{2011}).
\btitle{Geographic constraints on social network groups}.
\bjournal{PLoS ONE}
\bvolume{6}
\bpages{e16939}.
\bid{doi={10.1371/journal.pone.0016939}, issn={1932-6203},
pmcid={3071679}, pmid={21483665}}
\end{barticle}
%
\bptok{imsref}%
\endbibitem

\bibitem[\protect\citeauthoryear{Papadopoulos
et~al.}{2012}]{Papadopoulos2012}
%
\begin{barticle}[author]
\bauthor{\bsnm{Papadopoulos},~\bfnm{Symeon}\binits{S.}},
\bauthor{\bsnm{Kompatsiaris},~\bfnm{Yiannis}\binits{Y.}},
\bauthor{\bsnm{Vakali},~\bfnm{Athena}\binits{A.}} \AND
\bauthor{\bsnm{Spyridonos},~\bfnm{Ploutarchos}\binits{P.}}
(\byear{2012}).
\btitle{Community detection in social media}.
\bjournal{Data Min. Knowl. Discov.}
\bvolume{24}
\bpages{515--554}.
\end{barticle}
%
\bptok{imsref}%
\endbibitem

\bibitem[\protect\citeauthoryear{Porter, Onnela and
Mucha}{2009}]{porter2009communities}
%
\begin{barticle}[mr]
\bauthor{\bsnm{Porter},~\bfnm{Mason~A.}\binits{M.~A.}},
\bauthor{\bsnm{Onnela},~\bfnm{Jukka-Pekka}\binits{J.-P.}} \AND
\bauthor{\bsnm{Mucha},~\bfnm{Peter~J.}\binits{P.~J.}}
(\byear{2009}).
\btitle{Communities in networks}.
\bjournal{Notices Amer. Math. Soc.}
\bvolume{56}
\bpages{1082--1097}.
\bid{issn={0002-9920}, mr={2568495}}
\end{barticle}
%
\bptok{imsref}%
\endbibitem

\bibitem[\protect\citeauthoryear{Rosvall, Axelsson and
Bergstrom}{2009}]{rosvall2009map}
%
\begin{barticle}[author]
\bauthor{\bsnm{Rosvall},~\bfnm{Martin}\binits{M.}},
\bauthor{\bsnm{Axelsson},~\bfnm{Daniel}\binits{D.}} \AND
\bauthor{\bsnm{Bergstrom},~\bfnm{Carl~T.}\binits{C.~T.}}
(\byear{2009}).
\btitle{The map equation}.
\bjournal{The European Physical Journal Special Topics}
\bvolume{178}
\bpages{13--23}.
\end{barticle}
%
\bptok{imsref}%
\endbibitem

\bibitem[\protect\citeauthoryear{Rosvall and
Bergstrom}{2008}]{rosvall2008maps}
%
\begin{barticle}[author]
\bauthor{\bsnm{Rosvall},~\bfnm{M.}\binits{M.}} \AND
\bauthor{\bsnm{Bergstrom},~\bfnm{C.~T.}\binits{C.~T.}}
(\byear{2008}).
\btitle{Maps of random walks on complex networks reveal community structure}.
\bjournal{Proc. Natl. Acad. Sci. USA}
\bvolume{105}
\bpages{1118--1123}.
\end{barticle}
%
\bptok{imsref}%
\endbibitem

\bibitem[\protect\citeauthoryear{Rosvall and
Bergstrom}{2010}]{rosvall2010mapping}
%
\begin{barticle}[pbm]
\bauthor{\bsnm{Rosvall},~\bfnm{Martin}\binits{M.}} \AND
\bauthor{\bsnm{Bergstrom},~\bfnm{Carl~T.}\binits{C.~T.}}
(\byear{2010}).
\btitle{Mapping change in large networks}.
\bjournal{PLoS ONE}
\bvolume{5}
\bpages{e8694}.
\bid{doi={10.1371/journal.pone.0008694}, issn={1932-6203},
pmcid={2811724}, pmid={20111700}}
\end{barticle}
%
\bptok{imsref}%
\endbibitem

\bibitem[\protect\citeauthoryear{Shabalin et~al.}{2009}]{shabalin2009finding}
%
\begin{barticle}[mr]
\bauthor{\bsnm{Shabalin},~\bfnm{Andrey~A.}\binits{A.~A.}},
\bauthor{\bsnm{Weigman},~\bfnm{Victor~J.}\binits{V.~J.}},
\bauthor{\bsnm{Perou},~\bfnm{Charles~M.}\binits{C.~M.}} \AND
\bauthor{\bsnm{Nobel},~\bfnm{Andrew~B.}\binits{A.~B.}}
(\byear{2009}).
\btitle{Finding large average submatrices in high dimensional data}.
\bjournal{Ann. Appl. Stat.}
\bvolume{3}
\bpages{985--1012}.
\bid{doi={10.1214/09-AOAS239}, issn={1932-6157}, mr={2750383}}
\end{barticle}
%
\bptok{imsref}%
\endbibitem

\bibitem[\protect\citeauthoryear{Shi and Malik}{2000}]{shi2000normalized}
%
\begin{barticle}[author]
\bauthor{\bsnm{Shi},~\bfnm{J.}\binits{J.}} \AND
\bauthor{\bsnm{Malik},~\bfnm{J.}\binits{J.}}
(\byear{2000}).
\btitle{Normalized cuts and image segmentation}.
\bjournal{IEEE Transactions on Pattern Analysis and Machine Intelligence}
\bvolume{22}
\bpages{888--905}.
\end{barticle}
%
\bptok{imsref}%
\endbibitem

\bibitem[\protect\citeauthoryear{Snijders and
Nowicki}{1997}]{snijders1997estimation}
%
\begin{barticle}[mr]
\bauthor{\bsnm{Snijders},~\bfnm{Tom~A.~B.}\binits{T.~A.~B.}} \AND
\bauthor{\bsnm{Nowicki},~\bfnm{Krzysztof}\binits{K.}}
(\byear{1997}).
\btitle{Estimation and prediction for stochastic blockmodels for
graphs with latent block structure}.
\bjournal{J. Classification}
\bvolume{14}
\bpages{75--100}.
\bid{doi={10.1007/s003579900004}, issn={0176-4268}, mr={1449742}}
\end{barticle}
%
\bptok{imsref}%
\endbibitem

\bibitem[\protect\citeauthoryear{Traud, Mucha and
Porter}{2012}]{traud2012social}
%
\begin{barticle}[author]
\bauthor{\bsnm{Traud},~\bfnm{Amanda~L.}\binits{A.~L.}},
\bauthor{\bsnm{Mucha},~\bfnm{Peter~J.}\binits{P.~J.}} \AND
\bauthor{\bsnm{Porter},~\bfnm{Mason~A.}\binits{M.~A.}}
(\byear{2012}).
\btitle{Social structure of Facebook networks}.
\bjournal{Phys. A: Statistical Mechanics and Its Applications}
\bvolume{391}
\bpages{4165--4180}.
\end{barticle}
%
\bptok{imsref}%
\endbibitem

\bibitem[\protect\citeauthoryear{Traud et~al.}{2011}]{traud2011comparing}
%
\begin{barticle}[mr]
\bauthor{\bsnm{Traud},~\bfnm{Amanda~L.}\binits{A.~L.}},
\bauthor{\bsnm{Kelsic},~\bfnm{Eric~D.}\binits{E.~D.}},
\bauthor{\bsnm{Mucha},~\bfnm{Peter~J.}\binits{P.~J.}} \AND
\bauthor{\bsnm{Porter},~\bfnm{Mason~A.}\binits{M.~A.}}
(\byear{2011}).
\btitle{Comparing community structure to characteristics in online
collegiate social networks}.
\bjournal{SIAM Rev.}
\bvolume{53}
\bpages{526--543}.
\bid{doi={10.1137/080734315}, issn={0036-1445}, mr={2834086}}
\end{barticle}
%
\bptok{imsref}%
\endbibitem

\bibitem[\protect\citeauthoryear{Wei and Cheng}{1989}]{wei1989towards}
%
\begin{binproceedings}[author]
\bauthor{\bsnm{Wei},~\bfnm{Y.~C.}\binits{Y.~C.}} \AND
\bauthor{\bsnm{Cheng},~\bfnm{C.~K.}\binits{C.~K.}}
(\byear{1989}).
\btitle{Towards efficient hierarchical designs by ratio cut partitioning}.
In \bbooktitle{IEEE International Conference on Computer-Aided Design
(ICCAD-89). Digest of
Technical papers}
\bpages{298--301}.
\bpublisher{IEEE},
\blocation{New York}.
\end{binproceedings}
%
\bptok{imsref}%
\endbibitem

\bibitem[\protect\citeauthoryear{Wilson}{2014}]{supplemental}
%
\begin{bmisc}[author]
{\bauthor{\bsnm{Wilson},~\binits{J.}},
\bauthor{\bsnm{Wang},~\binits{S.}},
\bauthor{\bsnm{Mucha},~\binits{P.}},
\bauthor{\bsnm{Bhamidi},~\binits{S.}} \AND
\bauthor{\bsnm{Nobel},~\binits{A.}}}
(\byear{2014}).
\bhowpublished{Supplement to ``A testing based extraction algorithm
for identifying significant
communities in networks.''
DOI:\doiurl{10.1214/14-AOAS760SUPP}}.
\bptok{imsref}%
\end{bmisc}
%
\endbibitem


\bibitem[\protect\citeauthoryear{Xie, Kelley and
Szymanski}{2011}]{xie2011overlapping}
%
\begin{bmisc}[author]
\bauthor{\bsnm{Xie},~\bfnm{J.}\binits{J.}},
\bauthor{\bsnm{Kelley},~\bfnm{S.}\binits{S.}} \AND
\bauthor{\bsnm{Szymanski},~\bfnm{B.~K.}\binits{B.~K.}}
(\byear{2011}).
\bhowpublished{Overlapping community detection in networks: The state
of the art and comparative study.
Preprint. Available at \arxivurl{arXiv:1110.5813}.}
\end{bmisc}
%
\bptok{imsref}%
\endbibitem

\bibitem[\protect\citeauthoryear{Yang and
Leskovec}{2012}]{Yang2012Groundtruth}
%
\begin{binproceedings}[author]
\bauthor{\bsnm{Yang},~\bfnm{J.}\binits{J.}} \AND
\bauthor{\bsnm{Leskovec},~\bfnm{J.}\binits{J.}}
(\byear{2012}).
\btitle{Defining and Evaluating Network Communities based on Ground-truth}.
In \bbooktitle{Proceedings of the ACM SIGKDD Workshop on Data Semantics, 2012}.
\bpublisher{ACM},
\blocation{New York}.
\end{binproceedings}
%
\bptok{imsref}%
\endbibitem

\bibitem[\protect\citeauthoryear{Zhao, Levina and
Zhu}{2011}]{zhao2011community}
%
\begin{barticle}[author]
\bauthor{\bsnm{Zhao},~\bfnm{Y.}\binits{Y.}},
\bauthor{\bsnm{Levina},~\bfnm{E.}\binits{E.}} \AND
\bauthor{\bsnm{Zhu},~\bfnm{J.}\binits{J.}}
(\byear{2011}).
\btitle{Community extraction for social networks}.
\bjournal{Proc. Natl. Acad. Sci. USA}
\bvolume{108}
\bpages{7321--7326}.
\end{barticle}
%
\bptok{imsref}%
\endbibitem

\bibitem[\protect\citeauthoryear{Zhao, Levina and
Zhu}{2012}]{zhao2012consistency}
%
\begin{barticle}[mr]
\bauthor{\bsnm{Zhao},~\bfnm{Yunpeng}\binits{Y.}},
\bauthor{\bsnm{Levina},~\bfnm{Elizaveta}\binits{E.}} \AND
\bauthor{\bsnm{Zhu},~\bfnm{Ji}\binits{J.}}
(\byear{2012}).
\btitle{Consistency of community detection in networks under
degree-corrected stochastic block models}.
\bjournal{Ann. Statist.}
\bvolume{40}
\bpages{2266--2292}.
\bid{doi={10.1214/12-AOS1036}, issn={0090-5364}, mr={3059083}}
\end{barticle}
%
\bptok{imsref}%
\endbibitem
\end{thebibliography}
\end{document}